\documentclass[apj,numberedappendix]{emulateapj}
\usepackage{amsmath}

\shorttitle{Spectral properties of magnetohydrodnamic turbulence}
\shortauthors{Zhang, Lazarian \& Xiang}

\begin{document}
\title{Spectral properties of magnetohydrodnamic turbulence revealed by polarization synchrotron emission with Faraday rotation}

\author{Jian-Fu Zhang\altaffilmark{1,2}, Alex Lazarian\altaffilmark{3}, Fu-Yuan Xiang\altaffilmark{1} }
\altaffiltext{1}{Department of Physics, Xiangtan University, Xiangtan, Hunan 411105, China}\email{jfzhang@xtu.edu.cn(JFZ); alazarian@facstaff.wisc.edu(AL); fyxiang@xtu.edu.cn(FYX)}
\altaffiltext{2}{Key Laboratory for the Structure and Evolution of Celestial Objects,
Chinese Academy of Sciences, Kunming 650011, China}
\altaffiltext{3}{Astronomy Department, University of Wisconsin, Madison, WI 53711, USA}

\begin{abstract}
We investigate how to recover the spectral properties of underlying magnetohydrodynamic (MHD) turbulence using fluctuation statistics of synchrotron polarization radiation, based on the synthetic observations. Taking spatially coincident, separated, and compounded synchrotron emission and Faraday rotation regions into account, we extract the power spectrum of synchrotron polarization intensities integrated along the line of sight. Our results demonstrate that in the short wavelength range, the power spectra reflect fluctuation statistics of the perpendicular component of turbulent magnetic fields, and the spectra at long wavelengths reveal the fluctuation of the Faraday rotation density, which is a product of the parallel component of magnetic field and thermal electron density.  We find that our numerical results (in the case of spatially coincident regions) are in agreement with the analytical prediction in Lazarian \& Pogosyan (2016), and this theoretical prediction is applicable to more complicated settings, i.e., the spatially separated and compounded regions. We simulate telescopic observations that incorporate the effects of telescope angular resolution and noise, and find that statistics of underlying MHD turbulence can be recovered successfully. We expect that the technique can be applied to a variety of astrophysical environments, with existing synchrotron data cubes and a large number of forthcoming data sets from such as the LOw-Frequency Array for Radio astronomy (LOFAR), the Square Kilometer Array (SKA) and the Five-hundred-meter Aperture Spherical radio Telescope (FAST).
\end{abstract}

\section{Introduction}
\label{intro}
Both magnetic fields and turbulence are ubiquitous in astrophysical environments (\citealt{Elmegreen04,Scalo04,Ensslin06}). Turbulent motions perturb magnetic fields and result in magnetohydrodynamic (MHD) turbulence, which plays an important role in many key astrophysical processes, such as acceleration and propagation of cosmic rays (\citealt{Schlickeiser02}), star formation (\citealt{Elmegreen04,Mckee07,Lazarian12}), heat transfer (\citealt{Narayan01,Lazarian06}), and turbulent magnetic reconnection (\citealt{LV99,Kowal09,Kowal17,Lazarian15} for a review). In particular, the implications of MHD turbulence have been a focus of intensive studies in various astrophysical environments, such as Gamma-ray bursts (\citealt{ZhangB11,Lazarian18}), black hole X-ray binaries and active galactic nuclei (\citealt{de05,Kadowaki15,Zhang17,Zhang18}), as well as intracluster media (\citealt{Brunetti16}).

A significant progress has been achieved in understanding the theory of MHD turbulence (e.g., \citealt{Goldreich95,Cho03}; see also \citealt{BranL13} and \citealt{BereL15} for recent reviews), by using analytical, and especially numerical methods. However, it should be emphasized that the theory of MHD turbulence is a developing field with many open questions related to, such as the spectral slope of turbulence, the injection and dissipation of turbulence, the degree of turbulence compressibility, the degree of magnetization of turbulence. A proper understanding of these issues can shed light on important astrophysical processes. Therefore, a detailed study on quantifying the statistical properties of MHD turbulence is very necessary. Due to the limited resolution of current numerical simulations, it is difficult to obtain the realistic inertial range of astrophysical turbulence. New observational techniques provide more promising approaches for probing turbulence properties (\citealt{LP12,LP16}, henceforth LP12 and LP16).

Interactions of turbulent magnetic fields with non-thermal relativistic electrons produce synchrotron emissions carrying information on statistics of magnetic turbulence. Therefore, statistics of synchrotron emission can be used for studying MHD turbulence. LP12 provided a theoretical description of synchrotron intensity fluctuations arising from magnetic turbulence. Recently, \cite{Herron16} have tested their theoretical predictions by synthetic observations. Furthermore, it is well known that synchrotron emission is linearly polarized with the polarization direction perpendicular to the plane-of-sky magnetic field vector. When the polarized signal propagates through the magnetized plasmas, it can undergo depolarization in the presence of Faraday rotation fluctuations. The study incorporating both synchrotron emission and Faraday rotation (referred to as SEFR) was first presented in \cite{Burn66}, which, in general, is called traditional Faraday rotation synthesis. This method is commonly used in astrophysics and a remarkable progress has been made (e.g., \citealt{Brentjens05,Frick10,Frick11,Beck12,Akahori17}; see \citealt{Heald15} for a recent review).

By considering spatially coincident SEFR regions, LP16 proposed a new statistical description of synchrotron polarization fluctuations\footnote{On the basis of an analytical method, the case of the spatially separated SEFR regions has been discussed in Appendix C of LP16.}. Several measures of fluctuation statistics they proposed can be used to extract correlation scales and spectral slopes of the underlying MHD turbulence responsible for both SEFR fluctuations. Among them, two main measures are polarization spatial analysis (PSA), which makes use of spatial correlations of synchrotron polarization intensity (or its derivative) at the same wavelength as a function of the spatial separation $R$, and polarization frequency analysis (PFA), which considers a variance of synchrotron polarization intensity (or its derivative) as a function of the square of wavelength $\lambda^2$.

\cite{Zhang16} tested the PFA technique suggested in LP16 by using 1D, 2D, and 3D (three-dimensional) synthetic observations with high numerical resolution, and found that numerical results are in good agreement with the theoretical prediction of LP16. They stressed that the PFA technique can be used practically, because the effects of telescope's angular resolution and noise do not prevent the recovery of the underlying spectra of MHD turbulence. It is noticed that this technique has been applied to the depolarization of optical/infrared blazars (\citealt{Guo17}). In addition, using both synthetic and simulation data of MHD turbulence, \cite{Lee16} studied the power spectrum of synchrotron polarization intensity with \emph{zero mean} magnetic field. They demonstrated that simulations are in agreement with the theoretical prediction within the PSA technique of LP16.

\begin{figure*}[]
\centerline{\includegraphics[width=115mm,height=50mm,bb=260 270 890 500]{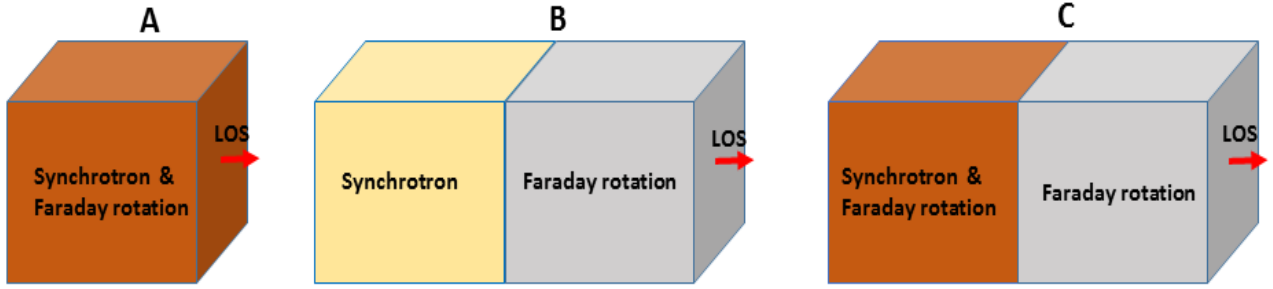}}
\caption{An illustration of spatial configurations: spatially coincident (panel A), separated (panel B), and compounded (panel C) synchrotron emission and Faraday rotation regions. Arrows represent the direction of the line of sight (LOS). }  \label{figs:THREESK}
\end{figure*}

As stressed above, intensive studies of synchrotron polarization statistics were focused on the scenario of the spatially coincident SEFR regions (LP12, LP16, \citealt{Herron16,Zhang16,Lee16}), which can be described explicitly from a mathematical point of view. However, in a real astrophysical environment, one has to confront with another situation that the intrinsic synchrotron polarization emission originates from one distinct region whereas Faraday rotation fluctuation occurs in another region. In addition, one could have to deal with a superposition of the above two limiting cases, which in this paper is termed spatially compounded SEFR regions (see Figure \ref{figs:THREESK} for an illustration).

The purpose of this paper, first of all, is to test theoretical prediction of the PSA technique of LP16 in the case of existence of \emph{non-zero mean} magnetic field. The second, mainly, is to study the more complicated setting compared to distributions of turbulence of relativistic electrons in LP16. We numerically study fluctuation statistics of synchrotron polarization emissions together with Faraday rotation measure (RM), from the spatially coincident, separated and compounded configurations. We focus on the influence of both direction and strength changes of mean magnetic fields, as well as the underlying spectral slope of MHD turbulence, on polarization intensity statistics. This work further paves stone for applying the new techniques suggested in LP16 to MHD turbulence on a wide variety of scales, such as star formation, supernova remnants, the Milky Way, distant galaxies and even clusters of galaxies. With existing synchrotron data sets and a large number of forthcoming data sets from the Very Long Array (VLA), and the Australian Square Kilometre Array Pathfinder (ASKAP), the Square Kilometer Array (SKA), the LOw Frequency ARray (LOFAR) and the Five-hundred-meter Aperture Spherical radio Telescope (FAST),
we expect that our techniques are very promising for understanding the origin and evolution of cosmic magnetism, which is a key science project of the SKA (\citealt{Beck15}).

The structure of the paper is as follows. In the next section, simulation methods for studying MHD turbulence are described. We present numerical tests of theoretical prediction of LP16 in Section 3, study more complicated spatial configurations for polarization SEFR regions in Section 4, and explore the influence of observational resolution and noise on the power spectra in Section 5. Discussion and summary are provided in Sections 6 and 7, respectively.

\section{Simulation methods}

\subsection{Synthetic data generation}\label{SynData}
The 3D synthetic data cubes for turbulent magnetic fields and thermal electron densities are generated in the wavenumber space ($k_x, k_y, k_z$), and are then transformed into a physical space. The generation of the former is governed by
\begin{equation}
\textbf{\textit{B}}(\textbf{\textit{r}})=\left<\textbf{\textit{B}}\right>+\sum_{k_{\rm min}\leq |\textbf{\textit{k}}|\leq k_{\rm max}}\widetilde{\textbf{\textit{B}}}(\textit{\textbf{k}})e^{i\chi} \label{Bxyz}
\end{equation}
with a condition of the solenoidal vector field $\triangledown\cdot \textit{\textbf{B}}=0$.  A periodic box of the size $2\pi$ in the Cartesian coordinate ($x, y, z$) is adopted, in which the $z$ axis is orientated along the line of sight. In Equation (\ref{Bxyz}), $\chi=\textbf{\textit{k}}\cdot \textit{\textbf{r}}+\vartheta(\textbf{\textit{k}})$ is the phase factor and  $\vartheta(\textbf{\textit{k}})$ is set as a random number between 0 and $2\pi$ with a periodic condition of $\vartheta(-\textbf{\textit{k}})=-\vartheta(\textbf{\textit{k}})$, where $\textbf{\textit{k}}=(k_x, k_y, k_z)$ and $\textbf{\textit{r}}=(x, y, z)$ are the wave vector and position vector, respectively. $\left<\textbf{\textit{B}}\right>$ is a mean magnetic field with components $\left<B_{x}\right>$, $\left<B_{y}\right>$ and $\left<B_{z}\right>$; the symbol $\left <...\right>$ denotes an average over the entire volume of interest. The Fourier coefficient $\widetilde{\textbf{\textit{B}}}(\textit{\textbf{k}})$ is a real vector that is perpendicular to $\textbf{\textit{k}}$ and whose squared amplitude follows a power-law scaling: $|\widetilde{\textbf{\textit{B}}}(\textit{\textbf{k}})|^2\propto k^{-\beta}$, $\beta$ representing the slope of magnetic turbulence, e.g., $\beta=11/3$ for Kolmogorov type. The generation procedure of data cubes for electron density is similar but no need for a divergence-free condition. The number density also includes the mean plus fluctuating components. The mean density is set to 1 and the root-mean-square of the fluctuation component is of the order of unity.

\subsection{Statistical description of turbulence}\label{StaDT}
MHD turbulence is a complex non-linear random process and exhibits some irregular features on a microscopic level. On the basis of macroscopic consideration, the statistical description is an effective way to shed light on regular features behind chaotic phenomena (\citealt{Biskamp03}). In practice, correlation and structure functions of any physical variable $\varPsi(\textbf{\textit{r}})$ have been traditionally used to characterize the properties of MHD turbulence. The correlation function is given by
\begin{equation}
{Corr}(\textbf{\textit{r}})=\left< \varPsi(\textbf{\textit{r}}_{\rm 1})\varPsi(\textbf{\textit{r}}_{\rm 2})\right>, \label{CorrF}
\end{equation}
where, $\textbf{\textit{r}}=\textbf{\textit{r}}_{\rm 2}-\textbf{\textit{r}}_{\rm 1}$ is a separation vector in 1D, 2D, or 3D spaces. As for the structure function, there is two- and multi-point second-order forms one can measure. In general, a two-point second-order scenario is commonly employed in the form of (but see \citealt{Cho09} for parallel direction of magnetic field)
\begin{equation}
{Stru}(\textbf{\textit{r}})=\left< [\varPsi(\textbf{\textit{r}}_{\rm 2})-\varPsi(\textbf{\textit{r}}_{\rm 1})]^2\right>=2\left [{Corr}(0)-{Corr}(\textbf{\textit{r}})\right ]. \label{StruF}
\end{equation}
As seen in Equation (\ref{StruF}), a two-point second-order structure function is related to the corresponding correlation function.

Furthermore, the power spectrum can also provide information concerning energy cascade processes in MHD turbulence, and is described as
\begin{equation}
P_{n\rm D}(\textbf{\textit{k}})={1\over (2\pi)^n}\int Corr(\textbf{\textit{r}}){\rm e}^{-i\textbf{\textit{k}}\cdot \textbf{\textit{r}}} d\textbf{\textit{r}}, \label{nDPS}
\end{equation}
via the Fourier transform of the correlation function. In Equation (\ref{nDPS}), $n=1, 2, 3$ denotes the number of the dimension of physical space.  On the basis of \cite{Lee16}, we in this paper adopt the following prescriptions for the power spectrum in 2D space, that is, the ring-integrated 1D spectrum for a 2D variable is given by
\begin{equation}
E_{\rm 2D}(k)=\int^{k+0.5}_{k-0.5}P_{\rm 2D}(k)dk, \label{E2DPS}
\end{equation}
where $P_{\rm 2D}=\varPsi(k)\varPsi^*(k)/2$ is a 2D power spectrum and the symbol `*' denotes the complex conjugate of $\varPsi(k)$. The integral of Equation (\ref{E2DPS}) is performed over a ring with inner ($k-0.5$) and outer ($k+0.5$) radii in 2D Fourier space.

There is a close connection between the correlation function, structure function, and power spectrum, as seen in Equations (\ref{CorrF}), (\ref{StruF}) and (\ref{nDPS}). Besides, the power-law indices for both the second-order structure function and the power spectrum are associated with the following relation. Namely, the structure function would follow $Stru(R)\propto R^{\alpha-1}$ if the power-law distribution of the ring-integrated 1D spectrum has $E_{\rm 2D}\propto k^{-\alpha}$  (see the appendix section of \citealt{Cho09} for more details).

\subsection{Polarization from synchrotron emission} \label{PFSE}
In this work, we assume that non-thermal relativistic electron population has a power-law energy distribution of $N(\varepsilon)\propto \varepsilon^{-p}$, where $p$ and $\varepsilon$ is the electron spectral index and energy, respectively. The spiralling motion electron would produce synchrotron polarization radiation in the environment of magnetic turbulence. We adopt the formulae of an isotropic pitch distribution to calculate the synchrotron polarization intensity (refer to Appendix A of LP16 for details). It is pretty obvious that the polarization intensity is proportional to $ |B_{\rm \perp}|^\gamma $, where  $\gamma=(p+1)/2$ and $B_{\rm \perp}=\sqrt{B_x^2+B_y^2}$ is the perpendicular component of magnetic field.

Generally, different combinations of the Stokes parameters can be chosen to describe the properties of synchrotron polarization. A complete description of the combinations, in the manner of encoding a polarization matrix, can be found in Appendix E of LP12. Alternatively, we in this paper focus on the linear polarization scenario that is characterized by the Stokes parameters $Q$ and $U$. Specifically, we consider the synchrotron polarization intensity as the form of $P=Q+iU$. In the plane of the sky, the synchrotron polarization intensity observed at a position vector $\textbf{\textit{X}}$ is expressed as
\begin{equation}
P(\textbf{\textit{X}},\lambda^2)=\int^{L_{\rm s2}}_{L_{\rm s1}}dzP_i(\textbf{\textit{X}},z){e}^{2i\lambda^2\varPhi(\textbf{\textit{X}},z)}, \label{PEq}
\end{equation}
where, $L_{\rm s2}$ and $L_{\rm s1}$ are the upper and lower boundaries of an emitting region, respectively. The integral is carried out along the line of sight $z$ over the emitting region. $P_i(\textbf{\textit{X}},z)$ denotes the intrinsic polarized intensity density, which is treated as wavelength-independent\footnote{In this paper, we also explore the wavelength dependence of $\lambda^{(p-1)/2}$, which arises from the distribution of relativistic electrons. We find that this wavelength dependence only changes the amplitude of the power spectrum and does not produce any variation of the slope.}, while the residual wavelength dependence is only from Faraday rotation. The intrinsic polarized emission is modified by the Faraday rotation of the polarization plane (e.g., \citealt{Burn66,Brentjens05}), which is reflected in the exponent factor of  ${e}^{2i\lambda^2\varPhi(\textbf{\textit{X}},z)}$.

When we study the scenario of the spatially coincident SEFR configuration (see Figure \ref{figs:THREESK}A), the Faraday RM is written as
 \begin{equation}
\varPhi(\textbf{\textit{X}},z)={e^3\over2\pi m_{\rm e}^2c^4}\int^z_{L_{\rm s1}}dz'n_{\rm e}(\textbf{\textit{X}},z')B_z(\textbf{\textit{X}},z'), \label{RM1}
\end{equation}
where $n_{\rm e}$ and $B_{z}$ are the number density of thermal electrons and the line of sight (parallel) component of the magnetic field, respectively. When we consider the spatially separated SEFR structure (see Figure \ref{figs:THREESK}B), Equation (\ref{RM1}) should be modified as
 \begin{equation}
\varPhi(\textbf{\textit{X}},L_{\rm f2})={e^3\over2\pi m_{\rm e}^2c^4}\int^{L_{\rm f2}}_{L_{\rm f1}}dz'n_{\rm e}(\textbf{\textit{X}},z')B_z(\textbf{\textit{X}},z'), \label{RM2}
\end{equation}
where ${L_{\rm f2}}$ and ${L_{\rm f1}}$ the boundaries of the external Faraday rotation region. Equations (\ref{RM1}) and (\ref{RM2}) should be combined correspondingly as we study the spatially compounded scenario (see Figure \ref{figs:THREESK}C). In other word, synchrotron emissions are subject to the effect of Faraday rotation within the emitting source, and they are also depolarized by another Faraday effect during the transfer process of the radiation. In our following simulations, we would use some typical parameters for the Galactic halo region, i.e., the depth of Faraday rotation of 100 pc, the electron density of 0.01 cm$^{-3}$, and the magnetic field strength of 1 $\rm \mu G$, to normalize the Faraday RM, $\varPhi$. For the calculation of the intrinsic polarization intensity, related physical quantities are calculated in arbitrary units for the sake of simplicity, because considering dimension does not change statistical results.

\subsection{Technique for simulating telescope observation}
We study the influence of angular resolution and random noise of the telescope on the ideal power spectrum, with a purpose of simulating a realistic observation. The main procedures we used are listed as follows.

1. Following the methods expressed in Section \ref{SynData}, we first generate 3D data cubes for magnetic fields (including $B_x$, $B_y$ and $B_z$) and the number density of thermal electrons $n_{\rm e}$. Then, we calculate the original polarized intensity `maps' related to Stokes $Q$ and $U$ according to Section \ref{PFSE} .

2. In order to study the influence of finite angular resolution, we convolve the original polarized intensity `maps' with a Gaussian kernel, which is given by
  \begin{equation}
g(x,y)={1\over \sigma\sqrt{2\pi}}e^{- {x^2+y^2\over2\sigma^2} }.
\end{equation}
The standard deviation of the Gaussian kernel is defined by $\sigma=\theta_{\rm FWHM}/2\sqrt{2{\rm ln}2}$ , in which $\theta_{\rm FWHM}$ is the full width at half maximum (FWHM) of the peak.

3. We generate a map of Gaussian noise and add then it into the original polarized intensity map to explore the influence of the noise. The level of the noise can be adjusted by changing the signal-to-noise ratio, which is defined as a fraction of the mean synchrotron intensity.

4. We mimic an interferometric observation, by randomly selecting wave vectors corresponding to the baselines of a telescope array in Fourier space (see \citealt{Lee16} for more details).

\section{Numerical tests of analytical prediction}\label{TESTPERD}
We first generate synthetic data cubes with 512 pixels along each direction in 3D space, by setting the scaling slopes of both turbulent magnetic field and density, as well as mean magnetic fields (see Section \ref{SynData}). Then, we calculate the maps of Stokes parameters $Q$ and $U$, which give rise to a map of the polarized intensity $P=Q+iU$. In this paper, we focus mainly on the statistical measurement of synchrotron polarization intensities by using the power spectrum.

In this section, we study polarization statistics from spatially coincident SEFR regions, that is, the synchrotron polarization radiation produced within the emitting-source region is subject to a Faraday rotation in the coincident space (see Figure \ref{figs:THREESK}A). We would test the influence of strength of mean magnetic field and its direction on the power spectrum and depolarization processes.

\begin{figure*}[]
\centerline{\includegraphics[width=70mm,height=95mm,angle=90]{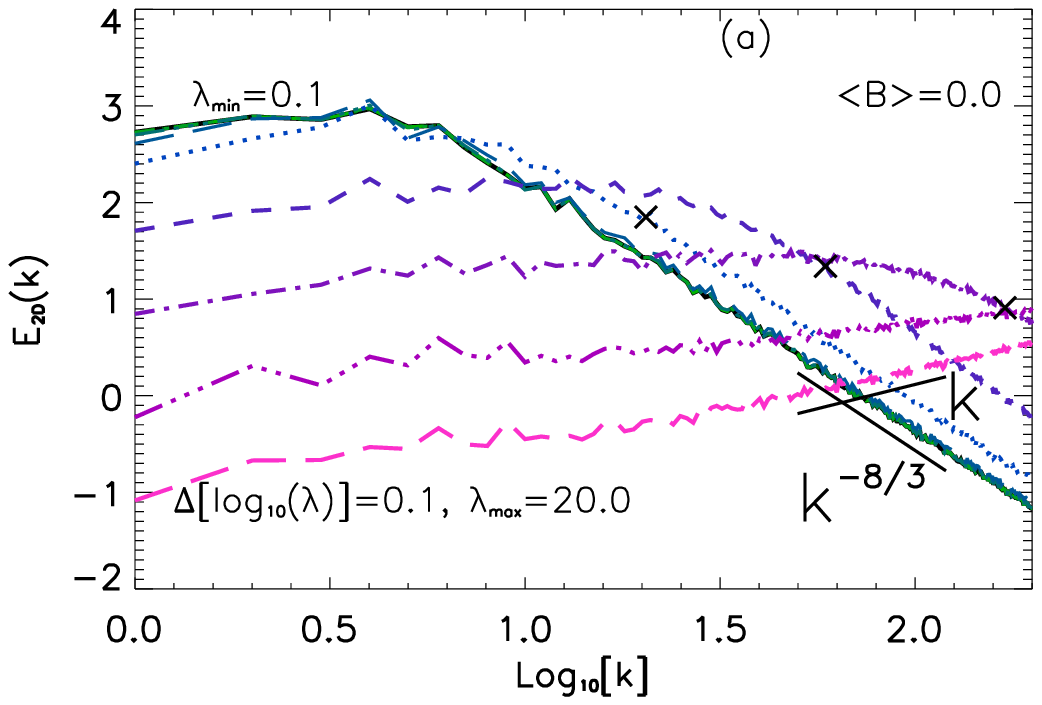} \includegraphics[width=70mm,height=95mm,angle=90]{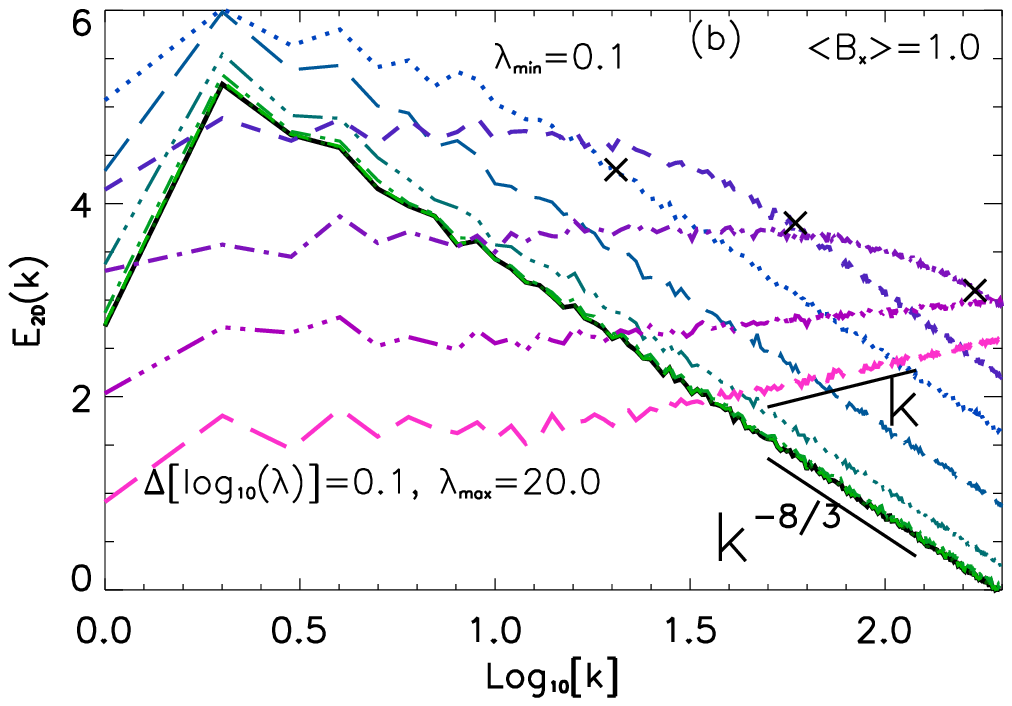}}
\centerline{\includegraphics[width=70mm,height=95mm,angle=90]{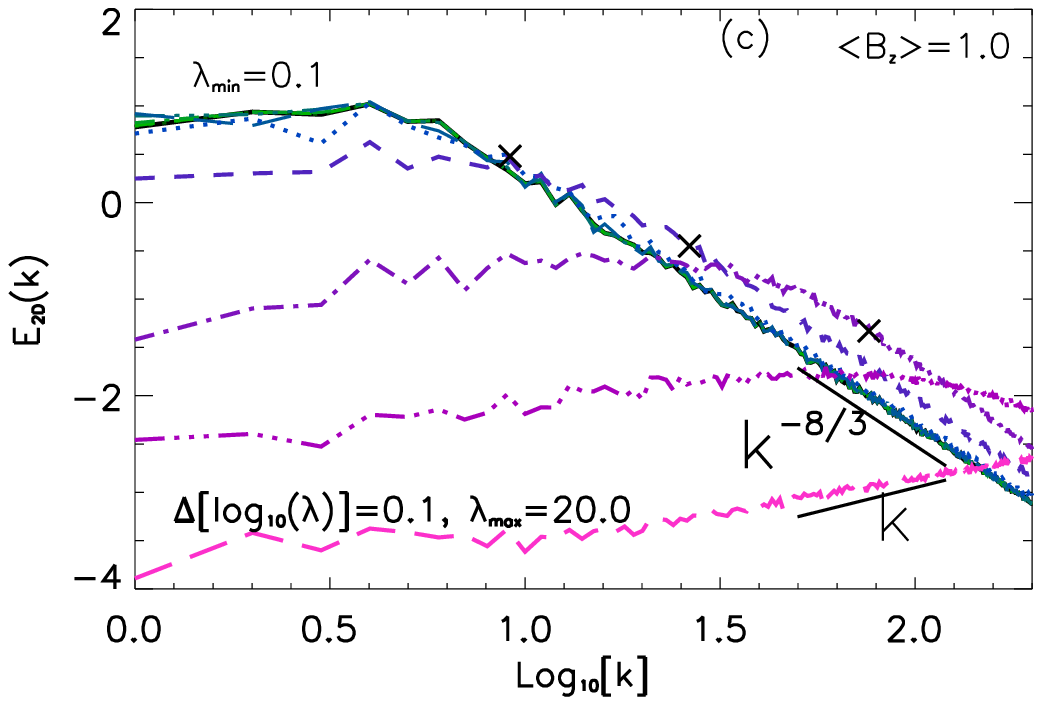} \includegraphics[width=70mm,height=95mm,angle=90]{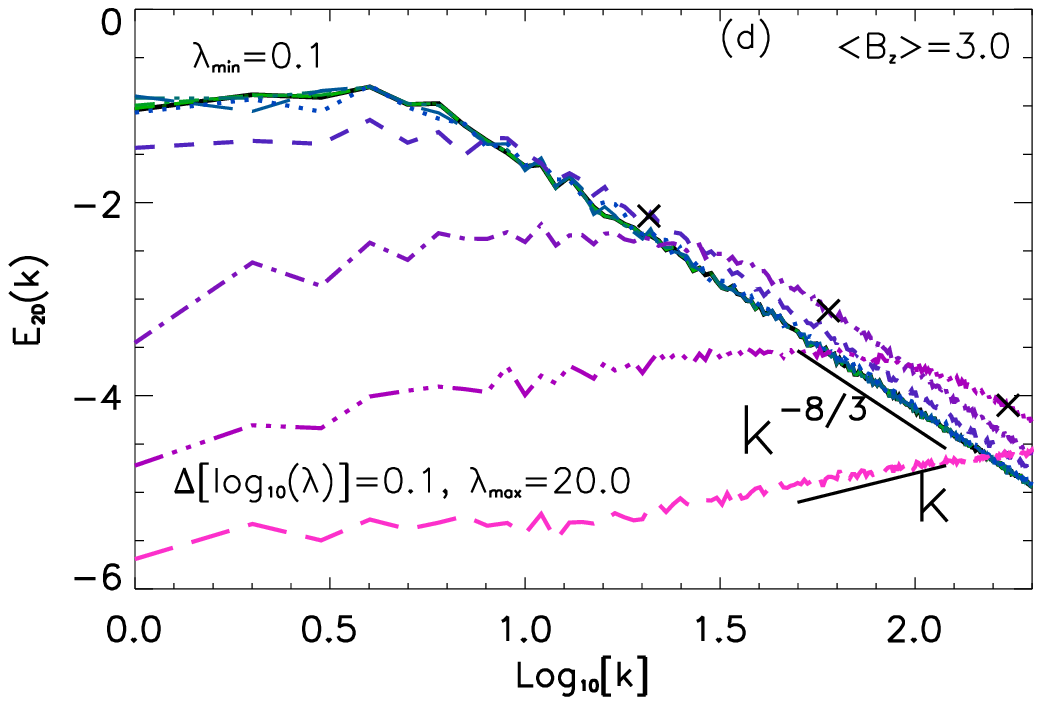}}
\caption{Ring-integrated 1D spectra $E_{\rm 2D}(k)$ of the synchrotron polarization intensity $P=Q+iU$, \emph{in the case of the spatially coincident SEFR regions}. Synthetic data cubes used are generated by setting the scaling slope of $\beta=11/3$ for both $B$ and $n_{\rm e}$, and different mean magnetic fields: $\left<B\right>=0$ for panel (a), $\left<B_{ x}\right>=1.0$ for panel (b), $\left<B_{z}\right>=1.0$ for panel (c) and $\left<B_{z}\right>=3.0$ for panel (d). The curves in each panel correspond to the wavelengths between $\lambda_{\rm min}=0.1$ (thick solid line) and $\lambda_{\rm max}=20.0$ in code units, in a logarithmic increment of $\bigtriangleup[{\rm log}_{10}(\lambda)]=0.1$. The symbols `X' indicate the wavenumber equal to $k=2\pi \lambda^2 \sigma_{\phi}$.
 }  \label{figs:COIN-PI}
\end{figure*}

Figure \ref{figs:COIN-PI} shows the ring-integrated 1D spectra $E_{\rm 2D}(k)$ of synchrotron polarization intensities for different mean magnetic fields but the same scaling slope of $\beta=11/3$ for both magnetic field and density. Power spectral distributions calculated at different wavelengths are plotted in each panel, ranging from the minimal wavelength $\lambda_{\rm min}=0.1$ in code units (thick solid line) to the maximal wavelength $\lambda_{\rm max}=20.0$, in a logarithmic increment of $\bigtriangleup[{\rm log}_{10}(\lambda)]=0.1$.

We first explain here the result for the case of zero mean magnetic field, i.e., $\left<B\right>=0$, as presented in Figure \ref{figs:COIN-PI}(a). It can be seen that at very short wavelengths Faraday rotation is negligible and power spectra ($\beta\sim8/3$) reflect the fluctuations of the perpendicular component of magnetic field. With increasing the wavelength, the amplitudes of power spectra in the small wavenumber $k$ range decrease due to the depolarization of Faraday rotation. However, they increase in the large-$k$ region because an additional fluctuation of the Faraday rotation density of $\phi=B_zn_{\rm e}$ manifests itself. In this paper, we adopt the definition of $k=2\pi \lambda^2 \sigma_{\phi}$ to characterize the degree of Faraday depolarization\footnote{It is noticed that \cite{Lee16} used the definition of $k=2\pi \lambda^2 \left<|B_z|n_{\rm e}\right>$ to describe the degree of Faraday depolarization, which can not work well for the case of non-zero mean fields studied in the current paper.}, where $\sigma_{\phi}$ and $2\pi$ represent the standard deviation of Faraday rotation density and the size of simulation box, respectively. The symbols `X' marked on each panel indicate the wavenumber equal to $k=2\pi \lambda^2 \sigma_{\phi}$ ($\sigma_{\phi}\sim 0.56$ for panel (a)). As shown, Faraday depolarization is significant at $k<2\pi \lambda^2 \sigma_{\phi}$ part of the wavenumber. However, this effect is insignificant at $k>2\pi \lambda^2 \sigma_{\phi}$ part, which issues fluctuation properties of RM density and provides a limit of measurement range of spectral scaling. In the range of $\lambda \ge \sqrt{k_{\rm max}/2\pi\sigma_{\phi}}$ ($k_{\rm max}$ is half of the maximum grid number), polarization radiation from any two space points gradually becomes uncorrelated, resulting in power spectra proportional to $k$.

Figure \ref{figs:COIN-PI}(b) presents the case of $\left<B_{ x}\right>=1.0$, i.e., the direction of mean magnetic field along the $x$-axis located in the plane of the sky. We see that amplitudes of power spectra significantly increase compared to that of panel (a). It is because that setting non-zero mean magnetic field $\left<B_{ x}\right>=1.0$ increases total strength of magnetic field $B_x$, which is mapped into the polarization intensity $P$ via $B_{\perp}=\sqrt{B_x^2+B_y^2}$. It seems that in the weak Faraday rotation region (i.e., short wavelengths) a measurable inertial range extends into more small-$k$ regime. Moreover, the amplitude of the power spectrum through the whole $k$ range increases with the wavelength, which should be due to the reason that the fluctuation of Faraday RM dominates that of the perpendicular component of magnetic field. In the long wavelength range, the Faraday depolarization dominates, resulting in a reduction of the power in the small-$k$ region. Likewise, the spectral slope of Faraday RM fluctuation can be recovered in the $k>2\pi \lambda^2\sigma_{\phi}$ regime.

The ring-integrated 1D spectra for the case of non-zero mean magnetic fields along the line of sight are explored in lower panels of Figure \ref{figs:COIN-PI}: $\left<B_{z}\right>=1.0$ for panel (c) and $\left<B_{z}\right>=3.0$ for panel (d), which correspond to $\sigma_{\phi}\sim 0.25$ and 0.20, respectively. It is understandable that fluctuation level of magnetic field decreases with increasing mean magnetic field, which gives rise to a small value of $k=2\pi \lambda^2 \sigma_{\phi}$ at a fixed wavelength. As a result, the small-$k$ value would widen the measurable inertial range of magnetic turbulence and enhance the chance to determine the spectral slope of the turbulence. Compared to that of panel (a), the amplitude of the power spectra decreases with increasing mean fields. It is because that the non-zero mean field setup would bring about a lower level of polarization fluctuations.

We above consider that the direction of non-zero mean magnetic field is oriented along the $x$-axis or $z$-axis. The fluctuation statistics of polarization radiation with mean magnetic field along the $y$-axis direction would have the same results as that of the $x$-axis, because $B_y$ is involved in $B_{\perp}$ in the same way as $B_x$. However, the mean magnetic field may have other orientations in a real astrophysical environment. We thus explore here a more general case that the presence of a mean magnetic field $\left<B\right>$ is confined in the $x$-$z$ plane for simplicity, in which the angle subtended by the vector of mean magnetic field and the $z$-axis is labeled as $\theta$. As an example, we consider the scenarios with different orientations $\theta=0^\circ$, $30^\circ$, $50^\circ$, $70^\circ$ and $90^\circ$, using $\left<B\right>=1.0$ and $\beta=11/3$ (the power spectral distribution not shown in this paper). We find that the behavior of the ring-integrated 1D power spectra in the case of $\theta=30^\circ$ is similar to that of Figure \ref{figs:COIN-PI}(b). As for the cases of $\theta=50^\circ$ and $70^\circ$, the ring-integrated 1D power spectra have similar characteristics like those of panels (c) and (d) of Figure \ref{figs:COIN-PI}, respectively. For the cases of $\theta=0^\circ$ and $90^\circ$, the resulting power spectra are the same as those of panels (b) and (c) in Figure \ref{figs:COIN-PI}, respectively. As a result, the scaling slope in the large-$k$ regime is close to the expected value of $8/3$. Accordingly, our numerical results are consistent with the theoretical prediction provided in Equation (103) of LP16.

\section{Simulations for more complicated settings}\label{SIMUSTUDY}

\subsection{Statistics from spatially separated synchrotron emission and Faraday rotation regions}\label{SSSS}

\begin{figure*}[]
\centerline{\includegraphics[width=70mm,height=95mm,angle=90]{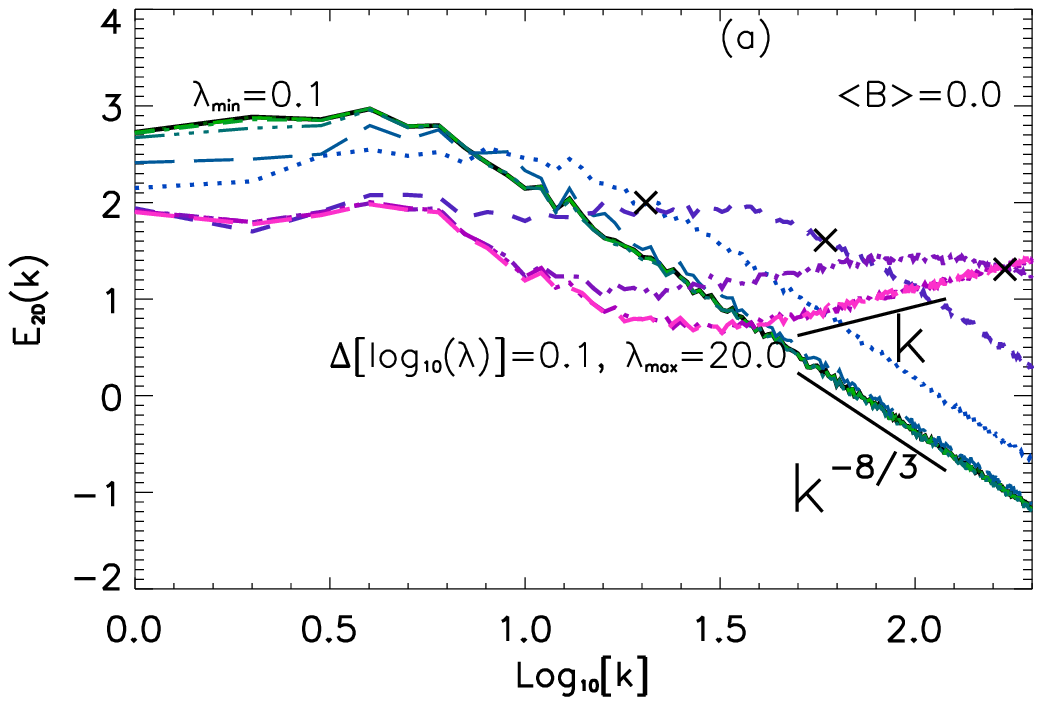} \includegraphics[width=70mm,height=95mm,angle=90]{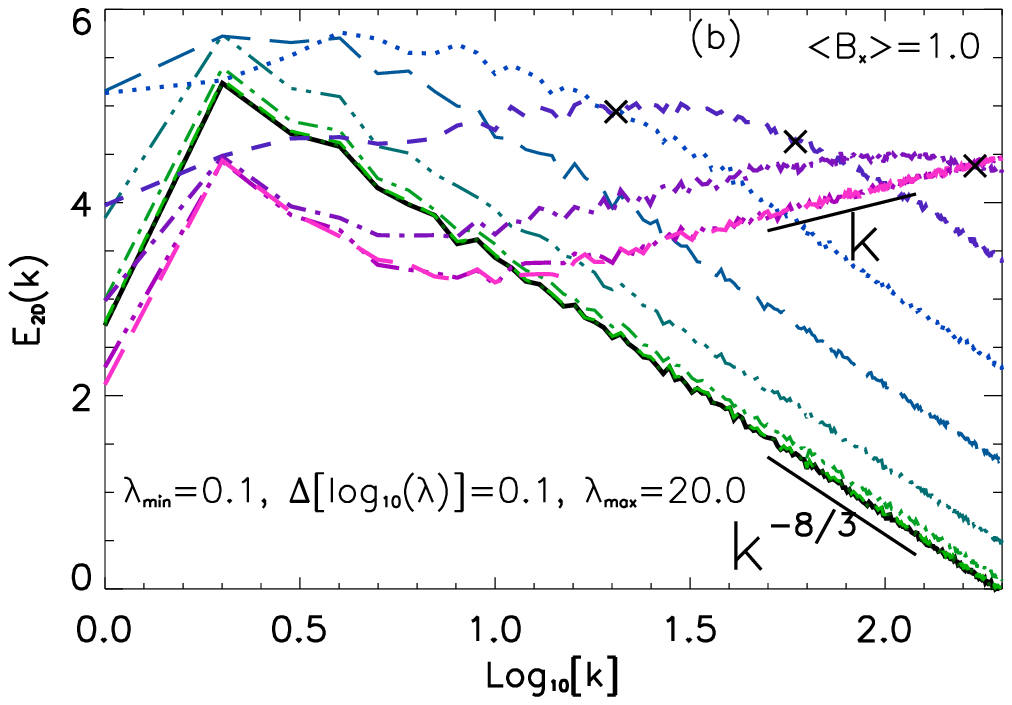}}
\centerline{\includegraphics[width=70mm,height=95mm,angle=90]{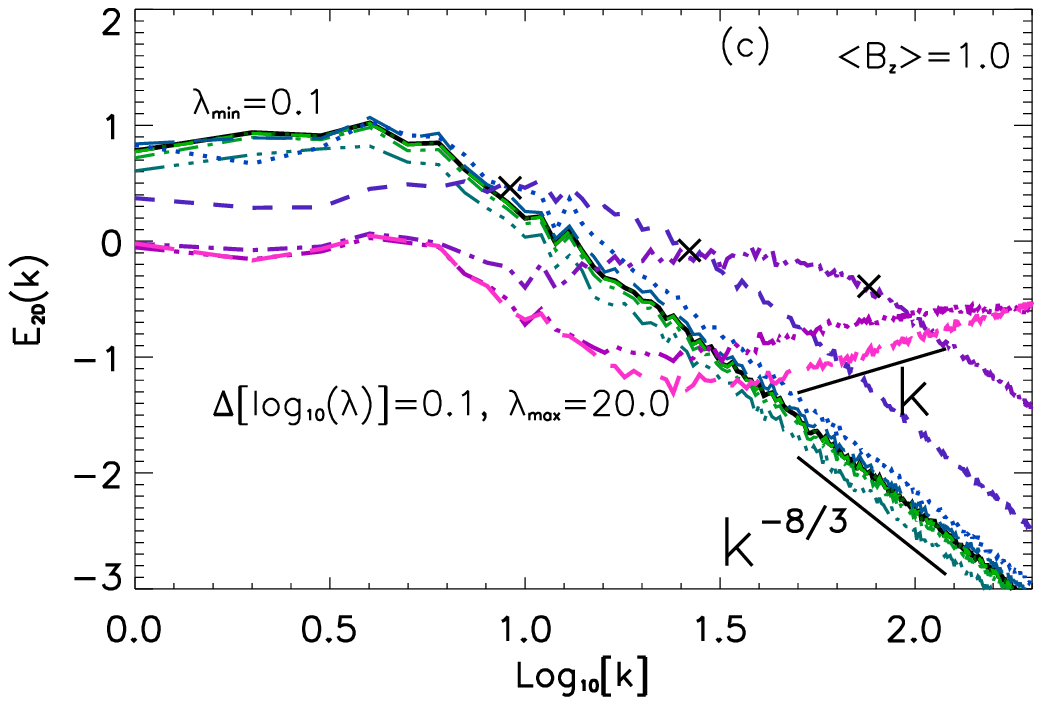} \includegraphics[width=70mm,height=95mm,angle=90]{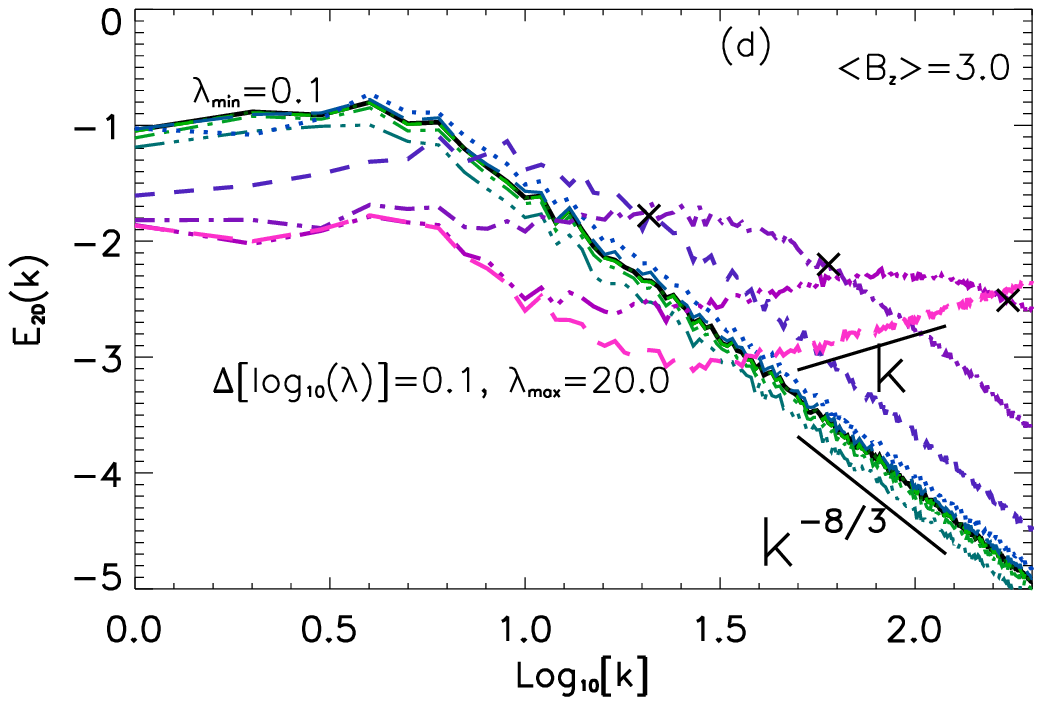}}
\caption{Ring-integrated 1D spectra $E_{\rm 2D}(k)$ of the synchrotron polarization intensity $P=Q+iU$, \emph{in the case of the spatially separated SEFR regions}. The other details are the same as for Figure \ref{figs:COIN-PI}.
 }  \label{figs:FRONT-PI}
\end{figure*}

In this section, we consider the case that intrinsic polarization synchrotron emission, i.e., without a Faraday rotation, is produced within the emitting source (termed background) while Faraday rotation is at work in another region (termed foreground). A possible example is the galactic environment, where polarized synchrotron emissions are emitted in the Galactic halo while depolarization is undergone in the region of the Galactic disk close to the Earth. We first generate synthetic data cubes using $\beta=11/3$ to simulate magnetic turbulence in the synchrotron emitting region. Moreover, the synthetic data cubes to mimic the turbulence in the Faraday rotation region are generated by using different seeds of random number.

The resulting ring-integrated 1D spectra $E_{\rm 2D}(k)$ of synchrotron polarization intensity are plotted in Figure \ref{figs:FRONT-PI} considering different mean magnetic fields along the $x$- and $z$-axes: $\left<B\right>=0$ for panel (a), $\left<B_{x}\right>=1.0$ (along the $x$-axis) for panel (b), $\left<B_{z}\right>=1.0$ for panel (c) and $\left<B_{z}\right>=3.0$ for panel (d) (along the $z$-axis). Spectral distributions shown in each panel correspond to the wavelengths ranging from $\lambda_{\rm min}=0.1$ to $\lambda_{\rm max}=20.0$, in a logarithmic increment of $\bigtriangleup[{\rm log}_{10}(\lambda)]=0.1$. As shown in panel (a), at very short wavelengths, the power spectrum reveals fluctuation properties of the perpendicular component of magnetic field in the background synchrotron emitting region, which has a slope close to 8/3. With increasing wavelengths, the information of the power spectrum reflects fluctuations of the Faraday rotation density in the foreground Faraday rotation region. At the small-$k$ part, Faraday depolarization makes the amplitude of the power decrease, while the power spectrum at the large-$k$ part, which presents the power-law of $k^{-8/3}$, increases due to raising fluctuation statistics of Faraday rotation density. In the longer wavelength range, the statistics of any two space points become uncorrelated.

As seen in Figure \ref{figs:FRONT-PI}(b), the non-zero mean magnetic field of $\left<B_{x}\right>=1.0$  make significant increase of the power (similar to panel (b) of Figure \ref{figs:COIN-PI}). However, for the non-zero mean magnetic field setup with direction along the $z$-axis, the amplitudes of the power spectra in the large-$k$ regime first decrease slightly on account of the depolarization, and increase then due to the dominated fluctuations of Faraday rotation density (see panels (c) and (d) of Figure \ref{figs:FRONT-PI}). In general, the reduction level in the amplitude of a power spectrum depends on the strength of mean magnetic fields along the line of sight.

Compared with Figure \ref{figs:COIN-PI}, i.e., for the spatially coincident case, a significant feature in Figure \ref{figs:FRONT-PI} is that between the wavenumber $k\simeq10$ and 100, the power spectrum at long wavelengths changes from fast decline to slow rise, forming a trough-like shape. Another feature is that the power reduction at small-$k$ part is not as dramatic as in the case of Figure \ref{figs:COIN-PI}. This is because of different extents of Faraday rotation interactions between spatially coincident and separated cases. As for the former, the polarization signal produced at different depths within the emitting source suffers from different degrees of Faraday rotation, that is, a distant signal is subject to more Faraday rotation effect, and vice versa. However, all intrinsic polarization  (background) signals have the same depth of Faraday rotation, in the case of the latter. Therefore, the power reduction at small-$k$ part depends on both wavelengths and differential Faraday rotation for the former, but only on wavelengths for the latter. In this section, we adopt the same integral length of Faraday rotation in the foreground medium (along the line of sight) as the scale of the background emission region for simplicity, which is normalized in our calculation by the typical parameters provided in Section \ref{PFSE}. Varying the extent of Faraday rotation region would not change the behavior of power spectra but can sample the polarized signals at different wavelengths. For instance, the deeper Faraday rotation region is, the larger the RM $\varPhi$ is, from which one can extract statistical information at shorter wavelengths.

\begin{figure*}[]
\centerline{\includegraphics[width=80mm,height=95mm,angle=90]{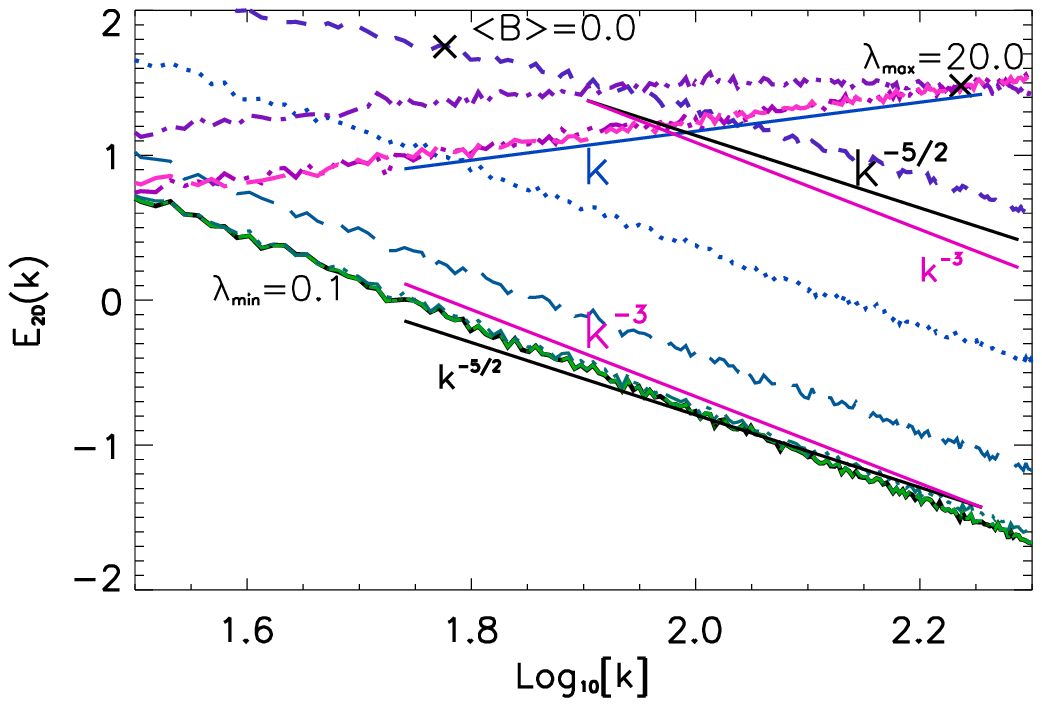} \includegraphics[width=80mm,height=95mm,angle=90]{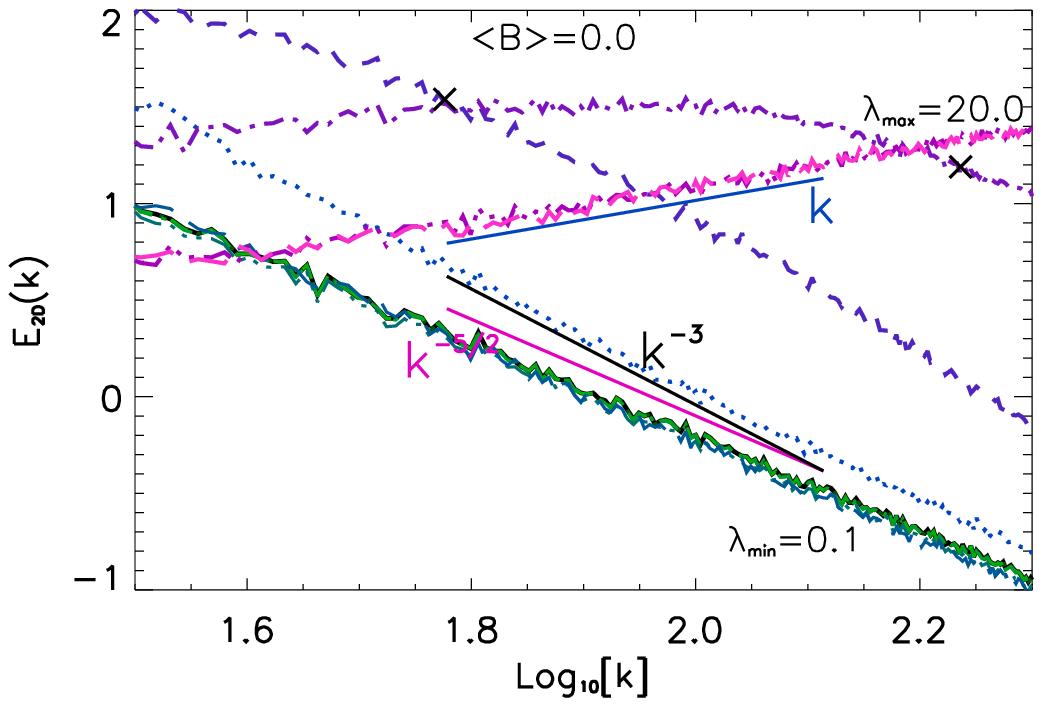}}
\caption{Ring-integrated 1D spectra $E_{\rm 2D}(k)$ of $P=Q+iU$, \emph{in the case of the spatially separated SEFR regions}. The parameters to generate synthetic data are provided as follows. \emph{Left panel}: $\beta=4$ for $B$ in the synchrotron-emitting region, $\beta=7/2$ for both $B$ and $n_{\rm e}$ in the Faraday rotation region. \emph{Right panel}: $\beta=7/2$ for $B$ in the synchrotron-emitting region, $\beta=4$ for both $B$ and $n_{\rm e}$ in the Faraday rotation region. The curves plotted correspond to the wavelength range between $\lambda_{\rm min}=0.1$ and $\lambda_{\rm max}=20.0$ in code units, in a logarithmic increment of $\bigtriangleup[{\rm log}_{10}(\lambda)]=0.1$. The symbols `X' denote the wavenumber to be $k=2\pi \lambda^2 \sigma_{\phi}$.
 }  \label{figs:DIFFFRONT}
\end{figure*}

\begin{figure}[]
\centerline{\includegraphics[width=80mm,height=95mm,angle=90]{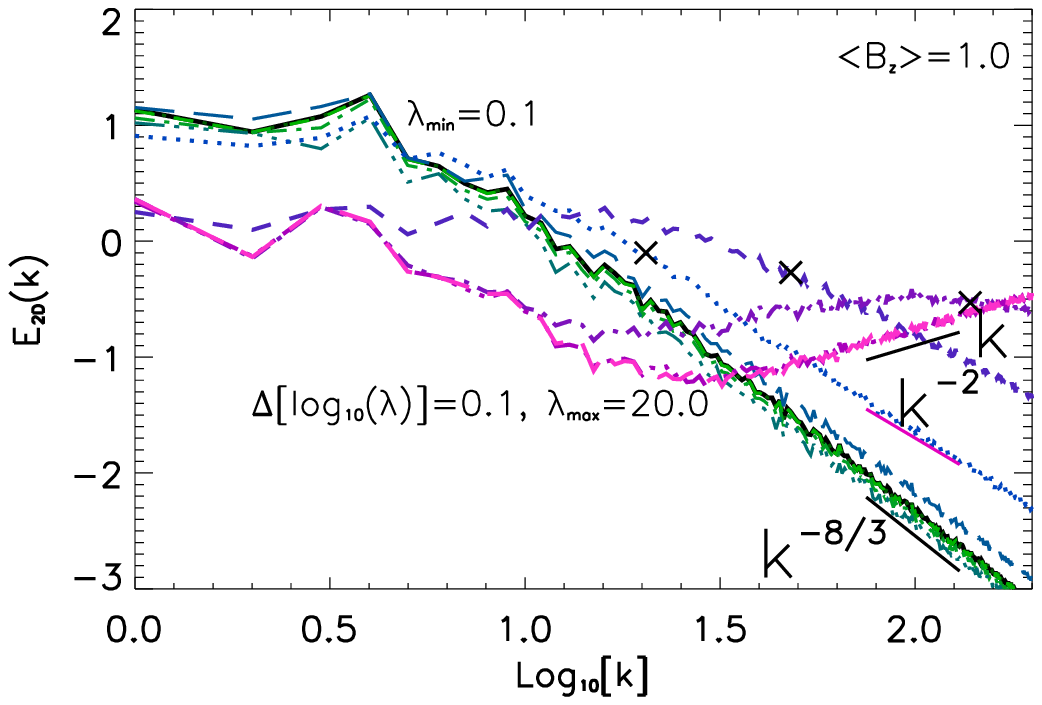}}
\caption{Ring-integrated 1D spectra $E_{\rm 2D}(k)$ of $P=Q+iU$, \emph{in the case of the spatially separated SEFR regions}. The parameters to generate synthetic data are given by $\beta=11/3$ for $B$ in the synchrotron-emitting region, $\beta=11/3$ for $B$ and $\beta=3$ for $n_{\rm e}$ in the foreground Faraday rotation region. The other descriptions are the same as for Figure \ref{figs:DIFFFRONT}.
 }  \label{figs:DIFFFRONT2}
\end{figure}

In order to distinguish fluctuation properties from the SEFR regions, we set different scaling slopes for both magnetic field and density to generate synthetic data. We first explore the case that both the magnetic field and density have a steep slope of $\beta=4$ in the synchrotron-emitting region but Kolmogorov spectrum of $\beta=7/2$ in the Faraday rotation region. We only present in Figure \ref{figs:DIFFFRONT} the power spectral distributions at large-$k$ part for clearly seeing differences between the slope with 3 and 5/2. It can be seen in the left panel that the power spectrum at very short wavelengths presents $E_{\rm 2D}(k)\propto k^{-3}$, which reflects the statistics of the pure synchrotron polarized fluctuations, i.e., without Faraday rotation effect, from the synchrotron emitting region. With increasing the wavelength, the contribution of Faraday rotation effect plays an increasingly important role. In the long wavelength range, the slope of the power spectrum becomes approximately $E_{\rm 2D}\propto k^{-5/2}$, which reveals fluctuations of both parallel component of magnetic field and density in the foreground space region. Furthermore, we swap synthetic data cubes used between synchrotron emission and Faraday rotation regions to calculate the power spectrum and show the results in the right panel. It is noticeable that spectra have the distributions of $E_{\rm 2D}\propto k^{-5/2}$ in very short wavelengths, and of $E_{\rm 2D} \propto k^{-3}$ in long wavelengths, which reflect the statistics of the perpendicular magnetic field component in the synchrotron emitting region, and of Faraday rotation density in the foreground space, respectively.

We here study how the scaling slope of the density influences the power spectrum of the polarization intensity. We consider that the density slope of $\beta=3$ is distinct from the magnetic field one (i.e., $\beta=11/3$), in the foreground space region. The power spectrum is plotted in Figure \ref{figs:DIFFFRONT2}. From this figure, we find that in the very short wavelength range spectra give $E_{\rm 2D}\propto k^{-8/3}$, which can recover well the statistics of fluctuations of the perpendicular component of the magnetic fields in the synchrotron emission region. With increasing the wavelength, the spectrum presents the distribution of $E_{\rm 2D}\propto k^{-2}$ in the large-$k$ range, which uncovers statistics of Faraday rotation density. It can be understood from the fact that the ring-averaged 1D spectrum $E_{\rm 2D}(k)$ of Faraday rotation density of $\phi=B_zn_{\rm e}$, where the slopes of $B_z$ and $n_{\rm e}$ are $\beta=11/3$ and $3$, respectively, follows the power-law distribution of $E_{\rm 2D}(k)\propto k^{-2}$.

\subsection{Statistics from spatially compounded synchrotron emission and Faraday rotation regions}\label{SSCS}

\begin{figure*}[]
\centerline{\includegraphics[width=70mm,height=95mm,angle=90]{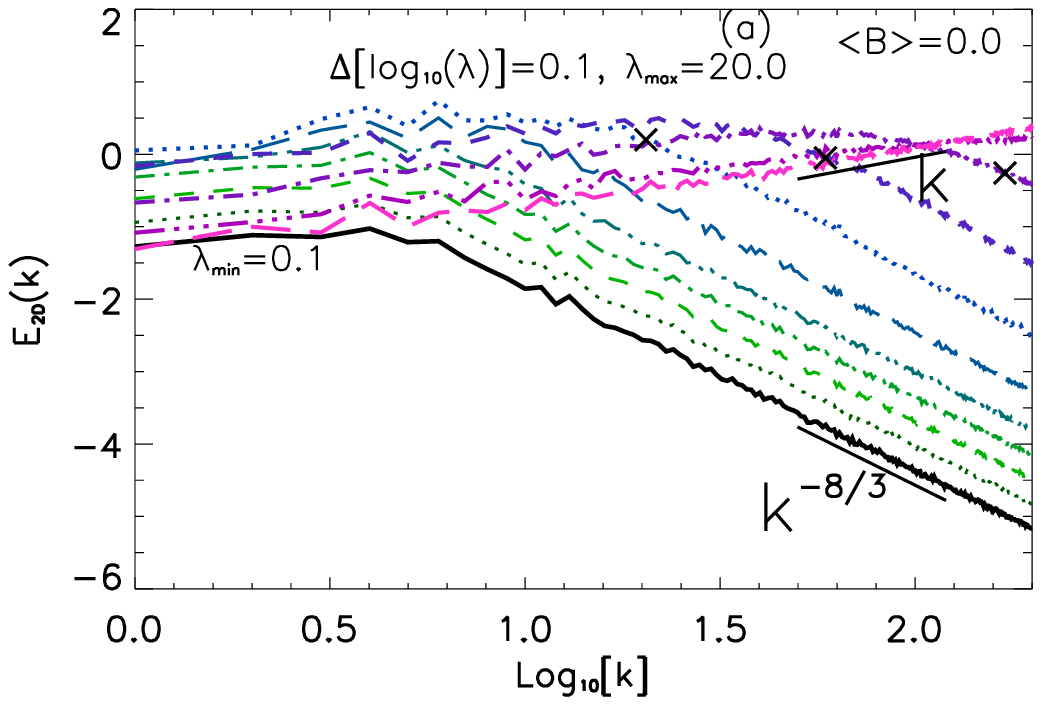} \includegraphics[width=70mm,height=95mm,angle=90]{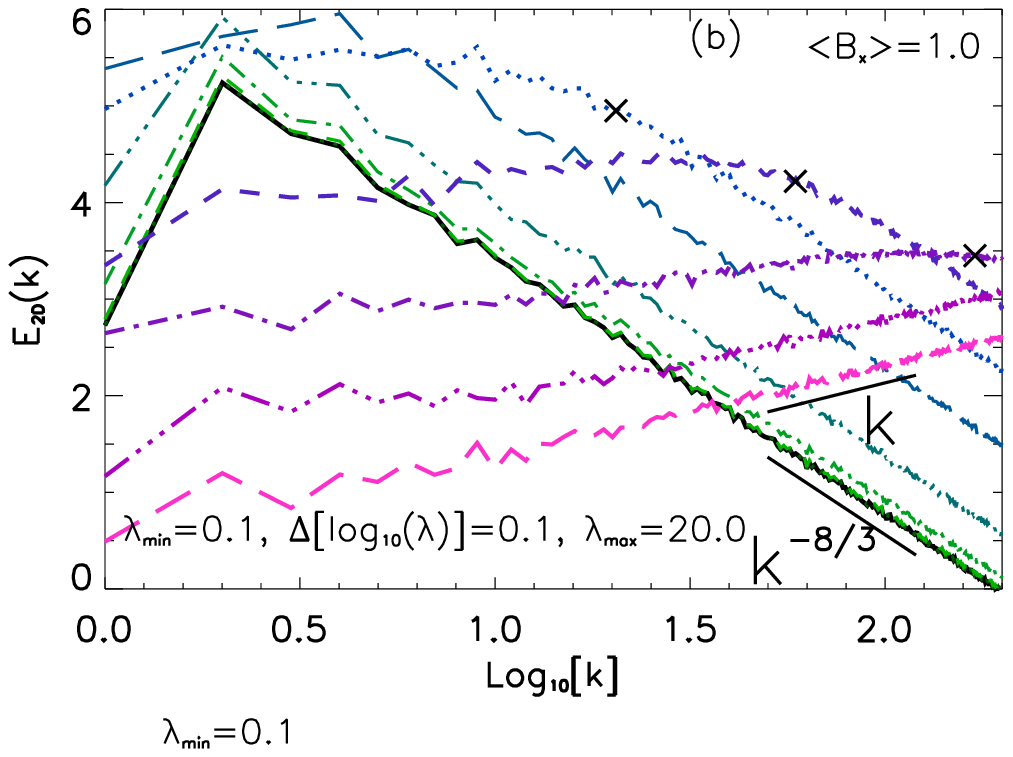}}
\centerline{\includegraphics[width=70mm,height=95mm,angle=90]{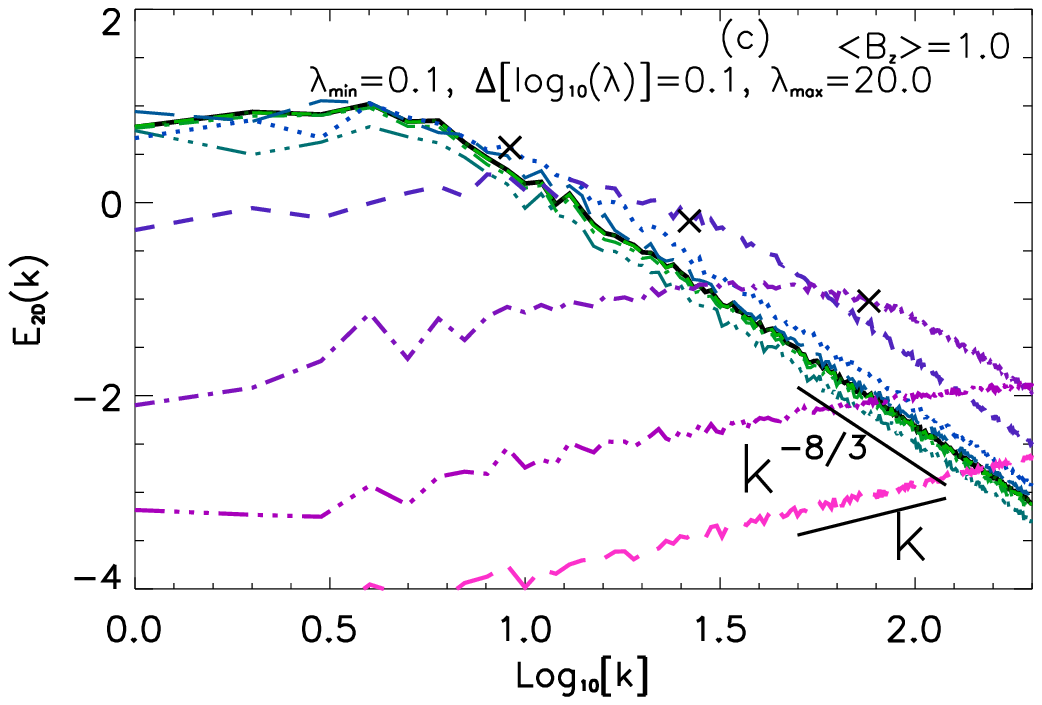} \includegraphics[width=70mm,height=95mm,angle=90]{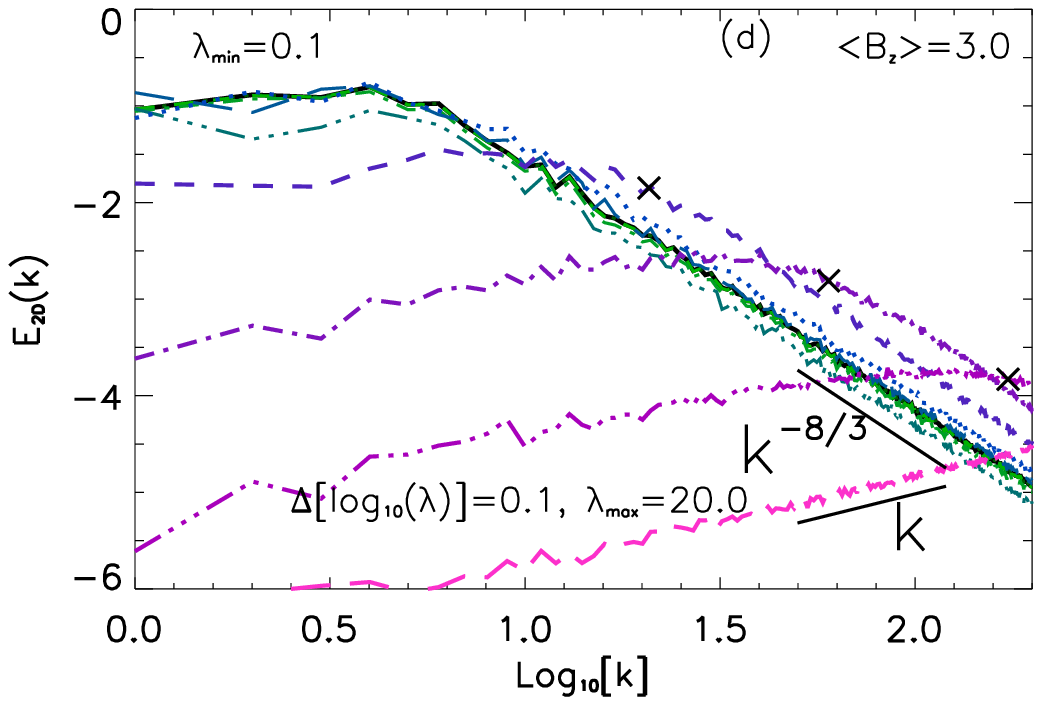}}
\caption{Ring-integrated 1D spectra $E_{\rm 2D}(k)$ of the synchrotron polarization intensity $P=Q+iU$, \emph{in the case of the compounded SEFR regions}. The other details are the same as those of Figure \ref{figs:COIN-PI}.
 }  \label{figs:TWOR-PI}
\end{figure*}

In a specific study of magnetic turbulence, one may confront with a combination of the two cases described above in Sections \ref{TESTPERD} and \ref{SSSS}. The background polarized synchrotron emissions undergo a Faraday rotation (similar to Section \ref{TESTPERD}), i.e., having a wavelength-dependence. Those polarized emissions are re-depolarized by another external Faraday rotation when they pass through a foreground medium (see Figure \ref{figs:THREESK}C).

In the case of the spatially compounded scenario, we study the influence of mean magnetic field on the power spectrum in Figure \ref{figs:TWOR-PI}. The result for the $\left<B\right>=0$ case is plotted in Figure \ref{figs:TWOR-PI}(a). It can be found that with increasing the wavelength, the amplitude of the power spectrum (including also the small-$k$ part) gradually become large, but the power-law part (i.e., so called inertial range) with a slope close to 8/3 gradually narrows down. At very short wavelengths, the power spectrum reflects fluctuations of intrinsic polarized synchrotron emission within the background space. When the wavelength increases, the contribution of fluctuations of Faraday rotation densities from the both background and foreground regions increases statistics of polarization radiation, which results in a continuous increase of the power. At longer wavelengths, the power reduction at small-$k$ part is due to the dominated Faraday depolarization effect.

Panels (b), (c) and (d) of Figure \ref{figs:TWOR-PI} show the results for the non-zero mean magnetic field of $\left<B_{x}\right>=1.0$, $\left<B_{z}\right>=1.0$ and $\left<B_{z}\right>=3.0$, respectively. From these panels, we find that the general behavior of the spectra is similar to that of Figure \ref{figs:COIN-PI} except the amplitude of the power spectrum slightly increases due to an additional fluctuation from the Faraday rotation density in the foreground space.

\begin{figure*}[]
\centerline{\includegraphics[width=80mm,height=95mm,angle=90]{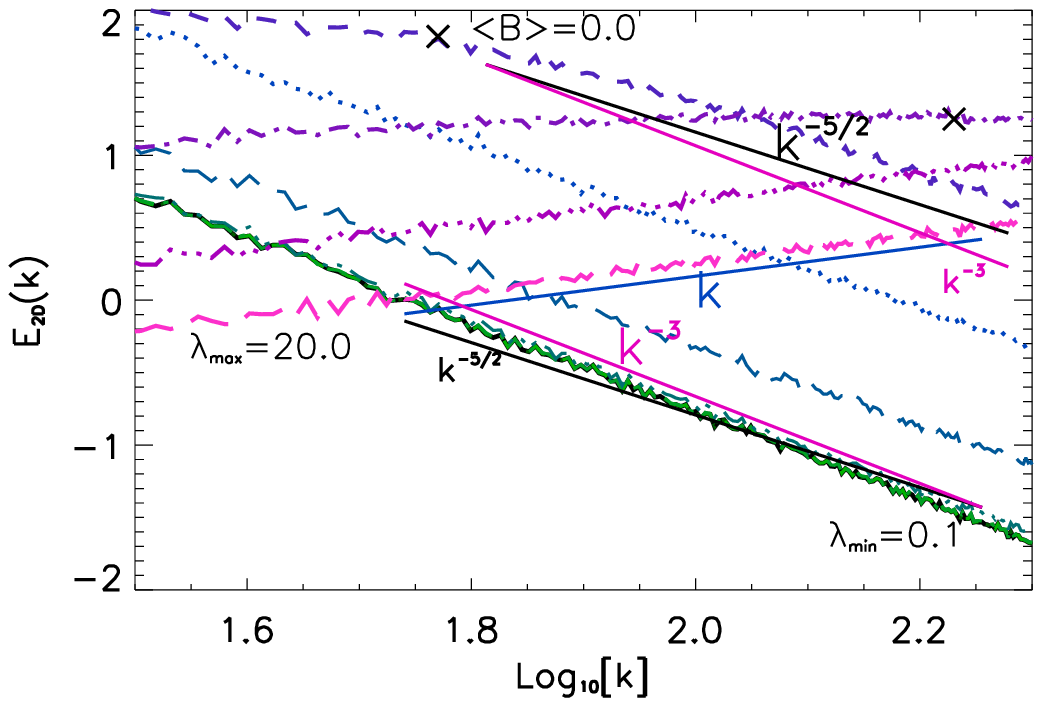} \includegraphics[width=80mm,height=95mm,angle=90]{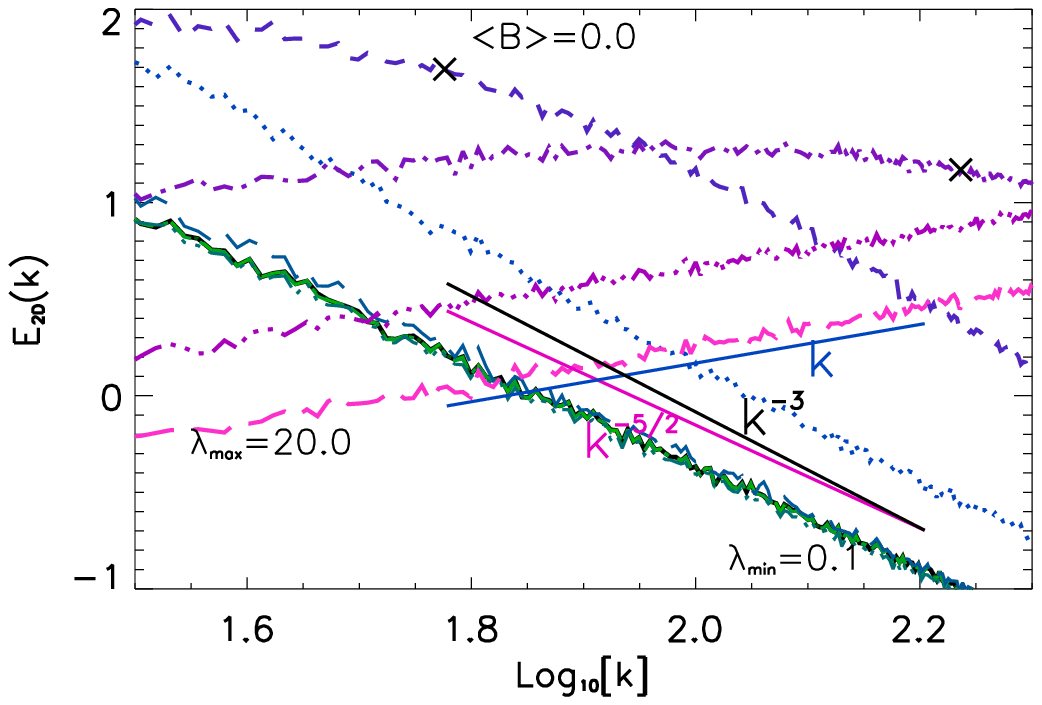}}
\caption{Ring-integrated 1D spectra $E_{\rm 2D}(k)$ of $P=Q+iU$, \emph{in the case of the compounded SEFR regions}. The used parameters to generate synthetic data cubes are given as follows. \emph{Left panel}: $\beta=4$ for both $B$ and $n_{\rm e}$ in the synchrotron-emitting region, $\beta=7/2$ for both $B$ and $n_{\rm e}$ in the foreground Faraday rotation region. \emph{Right panel}: $\beta=7/2$ for both $B$ and $n_{\rm e}$ in the synchrotron-emitting region, $\beta=4$ for both $B$ and $n_{\rm e}$ in the foreground Faraday rotation region. The other descriptions are the same as for Figure \ref{figs:DIFFFRONT}. }  \label{figs:DIFFMIXED}
\end{figure*}

\begin{figure}[]
\centerline{\includegraphics[width=80mm,height=95mm,angle=90]{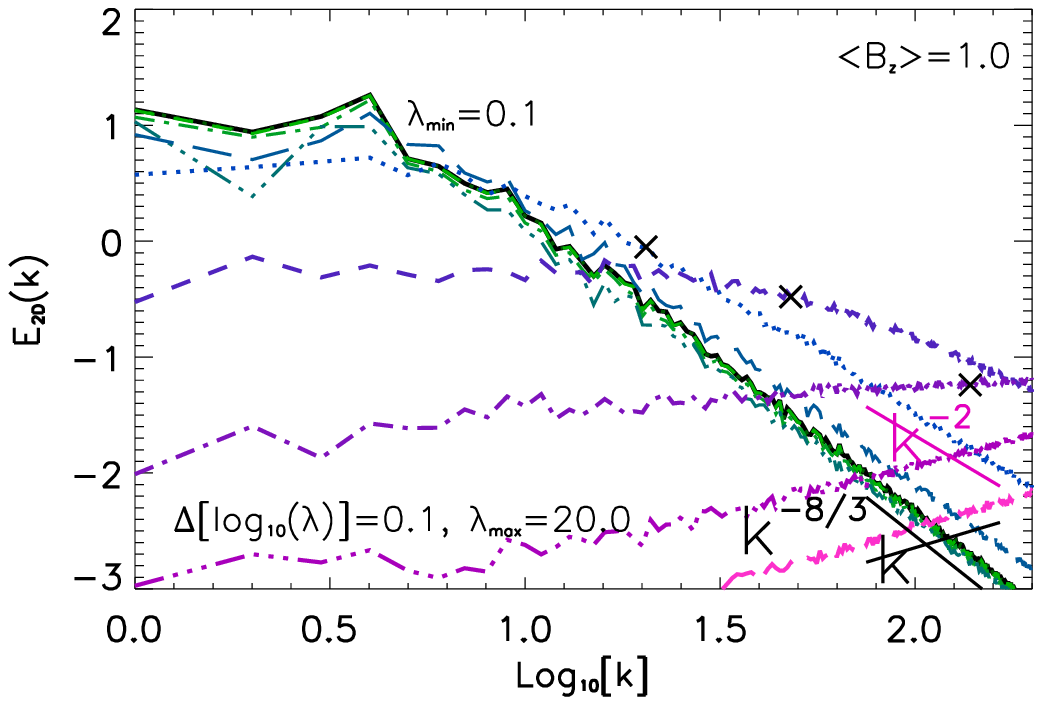}}
\caption{Ring-integrated 1D spectra $E_{\rm 2D}(k)$ of $P=Q+iU$, \emph{in the case of the compounded SEFR regions}. The used parameters to generate synthetic data cubes are given as follows: $\beta=11/3$ for both $B$ and $n_{\rm e}$ in the synchrotron-emitting region; $\beta=11/3$ for $B$ and $\beta=3$ for $n_{\rm e}$ in the foreground Faraday rotation region. The other descriptions are the same as for Figure \ref{figs:DIFFFRONT}. }  \label{figs:DIFFMIXED2}
\end{figure}

We then consider that spectral indices of synthetic magnetic field and density have a change in different Faraday rotation regions, in order to distinguish fluctuation statistics of two parts of Faraday rotation density. The slope of $\beta=4$ for both $B$ and $n_{\rm e}$ is used to generate synthetic data in the (background) synchrotron-emitting region having a Faraday effect, and $\beta=7/2$ for both $B$ and $n_{\rm e}$ in the foreground space. The ring-integrated 1D spectrum plotted in the left panel of Figure \ref{figs:DIFFMIXED} shows that Faraday depolarization makes the power spectrum itself offset intrinsic synchrotron polarization one (i.e., a slope close to $3$), which reflect the statistics of the Faraday rotation density fluctuation from the background region. With increasing the wavelength, the scaling slope would turn into $5/2$, which exposes Faraday fluctuation statistics from the foreground region. Furthermore, we explore an opposite situation in the right panel of Figure \ref{figs:DIFFMIXED}. The results show the spectrum ($E_{\rm 2D}\propto k^{-3}$ ) in the long wavelength range reveals the Faraday rotation statistics from the foreground space while the power spectrum ($E_{\rm 2D}\propto k^{-5/2}$) in the short wavelength range reflects polarization statistics from the background space region.

We next explore how the power spectrum is affected by the turbulent density. The parameters used to generate synthetic data cubes are given by: $\beta=11/3$ for $B$ and $\beta=3$ for $n_{\rm e}$ in the synchrotron-emitting region, which results in a ring-integrated 1D spectrum of $E_{\rm 2D}\propto k^{-8/3}$ of the Faraday rotation density $\phi_{\rm B}$, but $\beta=11/3$ for $B$ and $n_{\rm e}$ in the foreground Faraday rotation region, which gives $E_{\rm 2D}\propto k^{-2}$ of the Faraday rotation density $\phi_{\rm F}$. The result we obtained is plotted in Figure \ref{figs:DIFFMIXED2}, from which we find that in the short wavelength range the depolarization reveals the statistics of $\phi_{\rm B}$, with the scaling slope close to 8/3, while in the long wavelength range the depolarization reflects the statistics of $\phi_{\rm F}$, with the slope close to 2.

In brief, we can obtain statistics of magnetic fields, density, and Faraday rotation density, by selecting different wavelengths. Three scenarios we have studied are summarized in Table \ref{Table:SUM}. Although the complicated spatial configurations explored in this section cannot be described explicitly from an analytical study point of view, our simulations demonstrate that the general behavior of power spectra is in agreement with the theoretical prediction of the PSA technique in LP16. It should be stressed that the original PSA technique was based on the consideration of the spatially coincident SEFR regions, which is tested in Section \ref{TESTPERD} and \cite{Lee16}. As a result, the PSA technique of LP16 can be applied to these complicated settings.

\begin{table}
\caption{Studying how to extract MHD turbulence information at different wavelengths for three spatial configurations.}
\begin{tabular}{cccc}
\hline\hline
Cases & Very short $\lambda$ & Short $\lambda$ & Long $\lambda$  \\
\hline
 Coincident &  $B_{\rm \perp}$  & $\phi$, $B_{\rm \parallel}$, $n_{\rm e}$ & $\phi$, $B_{\rm \parallel}$, $n_{\rm e}$  \\
\hline
 Separated &  $B_{\rm \perp B}$  & $\phi_{\rm B}$, $B_{\rm \parallel B}$, $n_{\rm eB}$ & $\phi_{\rm F}$,$B_{\rm \parallel F}$, $n_{\rm eF}$  \\
 \hline
 Compounded &  $B_{\rm \perp B}$  & $\phi_{\rm B}$, $B_{\rm \parallel B}$, $n_{\rm eB}$ & $\phi_{\rm F}$,$B_{\rm \parallel F}$, $n_{\rm eF}$  \\
\hline\\
\end{tabular}
\\ \textbf{Note}. Symbol indicating $\lambda$: wavelength; $B_{\rm \perp}$: perpendicular component of magnetic field; $B_{\rm \parallel}$: parallel component of magnetic field; $\phi$: Faraday rotation density; $n_{\rm e}$: electron density. The subscript $\rm B$ and $\rm xB$ ($\rm x=\perp, \parallel$, or $\rm e$) correspond to physical quantities in the (background) synchrotron emitting region, and $\rm F$ and $\rm xF$ in the (foreground) Faraday rotation region.
\label{Table:SUM}
\end{table}

\section{Influence of observations on power spectrum}

\begin{figure*}[]
\centerline{\includegraphics[width=80mm,height=95mm,angle=90]{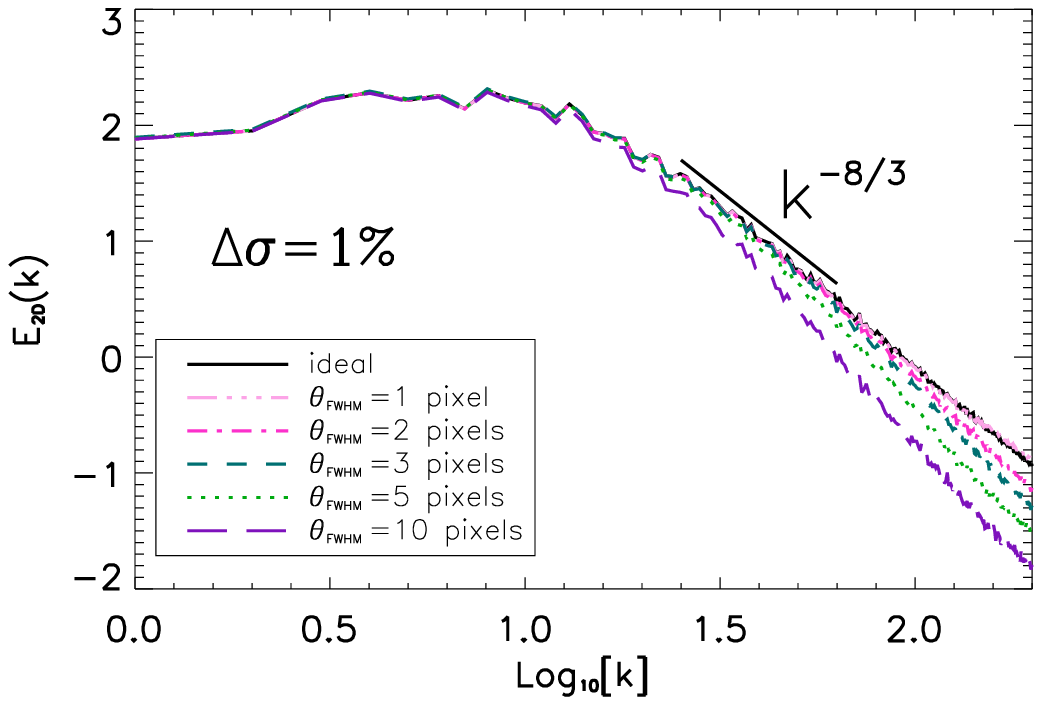} \includegraphics[width=80mm,height=95mm,angle=90]{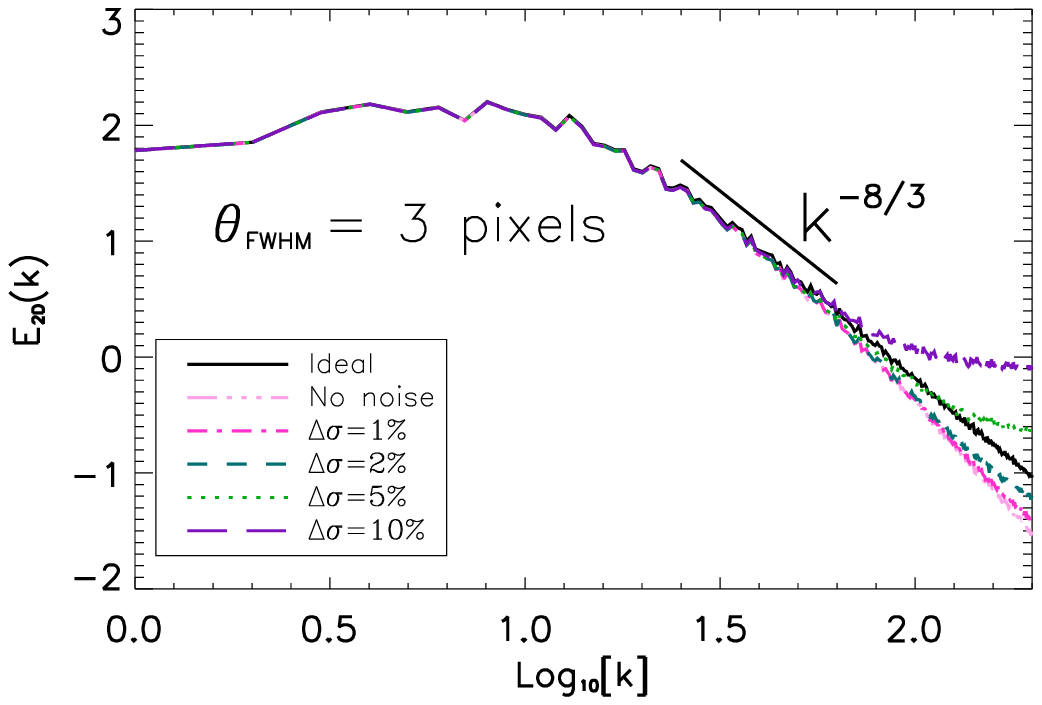}}
\caption{The influences of telescope's angular resolution (left panel) and Gaussian noise (right panel) on the turbulent spectrum calculated at the wavelength $\lambda=2.4$ in code units, with $\beta=11/3$ for $B$ and $n_{\rm e}$, \emph{in the case of the spatially separated SEFR regions}. The symbol $\Delta\sigma$ indicates standard deviation of Gaussian noise and accounts for a fraction of the mean synchrotron intensity. The symbol $\theta_{\rm FWHM}$ denotes an effective Gaussian beam, which is used to convolve 2D maps. The original (ideal) slope without considering the resolution and noise is shown in the solid line.
 }  \label{figs:ANG-NOISE-A}
\end{figure*}

\begin{figure}
        \centerline{\includegraphics[width=80mm,height=95mm,angle=90]{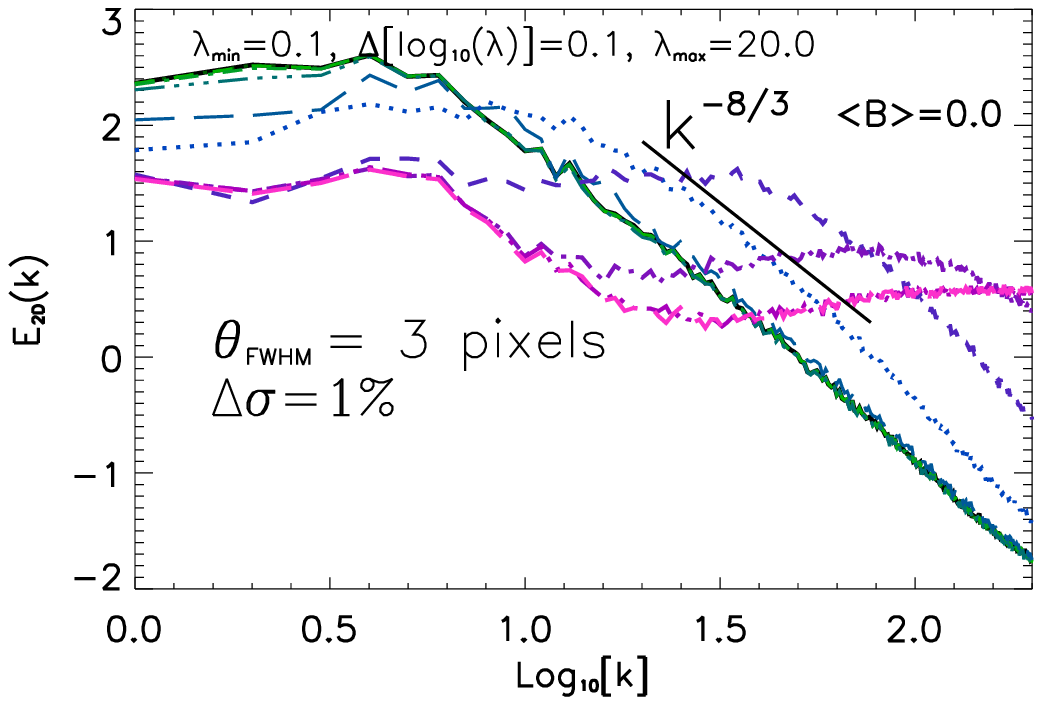}}
\caption[ ]{Ring-integrated 1D spectra $E_{\rm 2D}(k)$ of $P=Q+iU$ obtained considering the influence of both noise and resolution, \emph{in the case of the spatially separated SEFR regions}. The curves plotted in each panel correspond to the wavelength range between $\lambda_{\rm min}=0.1$ (thick solid line) and $\lambda_{\rm max}=20.0$ in code units, in a logarithmic increment of $\bigtriangleup[{\rm log}_{10}(\lambda)]=0.1$.
} \label{figs:ANG-NOISE-B}
\end{figure}

We study the influence of telescope's noise and angular resolution on the power spectrum of polarization intensity to simulate a realistic observation. As an example, we first obtain 2D maps for Stokes parameters $I$, $Q$ and $U$ (integrated along the line of sight $z$) at different wavelengths, in the case of the spatially separated SEFR regions, which has been explored in Section \ref{SSSS}. Those 2D maps are termed original (idealized) images with the size of $512^2$. With a Gaussian kernel, we then smooth original maps in an expected pixel that corresponds to the standard deviation of a telescope beam via $\sigma=\theta_{\rm FWHM}/2\sqrt{2{\rm ln}2}$. In this work, we adopt a code unit to perform simulations. When it is necessary, code units can be easily switched to real physical units. For instance, if one considers that one pixel corresponds to $0.01\ \rm pc$ and the distance of emitting source to the Earth is $2\  \rm kpc$, one has an angular distance of 1 arcsec for one pixel.

In order to mimic a telescopic noise, we also use a Gaussian kernel to generate a noise map with $512^2$ sizes, in which we consider a standard deviation equal to a fraction of mean synchrotron intensity, and add then the noise map to an original image. At the wavelength $\lambda=2.4$ in code units, we present an exemplification as plotted in Figure \ref{figs:ANG-NOISE-A}, considering angular resolution and noise effects. The left panel of the figure shows the influence of different angular resolutions $\theta_{\rm FWHM}$ on the ring-integrated 1D spectrum, and the right panel presents the influence of different noise levels $\Delta\sigma$ on the spectrum. We find that with decreasing angular resolutions, i.e., increasing the value of $\theta_{\rm FWHM}$, ring-integrated 1D spectra gradually deviate downwards from the ideal slope of $8/3$ and become more steep, in particular, at large-$k$ part (see left panel). We know that smoothing processes cancel out small features of maps and reduce random statistics of synchrotron polarization intensity. On the contrary, the spectrum in the large-$k$ regime deviates upwards from the original spectrum, with increasing the level of Gaussian noise (see right panel). The added noise map equivalently increases random fluctuations that result in an increase of the power at large-$k$ part. Consequently, an optimal range to recover slope of MHD turbulence should range from $k\approx 15$ to 100.

\begin{figure}
        \centerline{\includegraphics[width=95mm,height=80mm,angle=0]{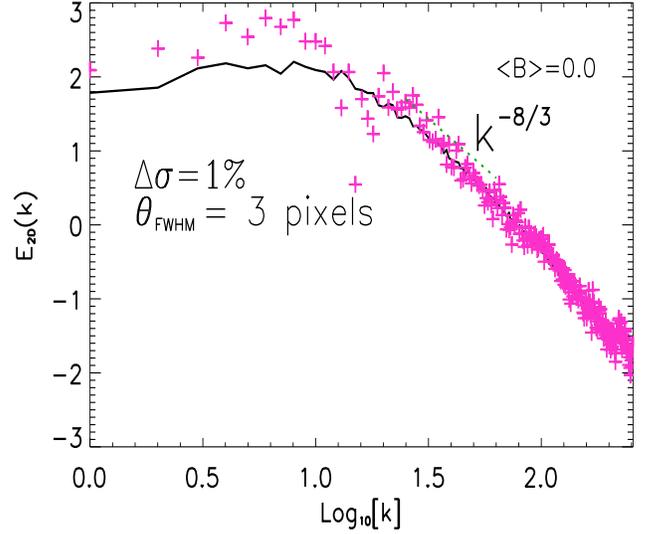}}
\caption[ ]{Ring-integrated 1D spectrum $E_{\rm 2D}(k)$ of $P=Q+iU$ from interferometric observations simulated by considering both noise and resolution, \emph{in the case of the spatially separated SEFR regions}. Plus symbols indicate the randomly selected wavenumber of 160 antenna arrays and the true spectrum is shown by the solid line. Simulations are performed at the wavelength $\lambda=2.4$ in code units, with the parameter of $\beta=11/3$ for $B$ and $n_{\rm e}$.
  } \label{figs:INTERFERO}
\end{figure}

Furthermore, we consider the influence of both angular resolution and Gaussian noise on the idealized spectra at different wavelengths. As an example, on the basis of the parameters used in Figure \ref{figs:FRONT-PI}(a), we first add a noise map with the noise level $\Delta\sigma=1\%$ to original maps. Then, we convolve new maps to $\theta_{\rm FWHM}=3$ pixels, because usual telescope surveys have a typical value of 3 pixels. As seen in Figure \ref{figs:ANG-NOISE-B}, the scaling slope of power spectra can be determined in the range of $k\gtrsim15$ to $k\simeq 100$.

One of advantages of new techniques proposed in LP16 is that one can directly adopt interferometric data to obtain the power spectral distribution of underlying MHD turbulence. By adopting a similar method as expressed in \cite{Lee16}, we here study how to use an interferometric observation to recover MHD turbulence spectrum. We assume that an interferometer consists of $N$ antennas that corresponds to $N_{\rm B}=N(N-1)/2$ baselines. We randomly select $N_{\rm B}$ wave vectors from the map of polarization intensity to calculate the new ring-integrated 1D spectra.  In Figure \ref{figs:INTERFERO}, we present the result simulated with 160 antennas at the wavelength $\lambda=2.4$ in code units. It is obvious that employing a certain amount of antenna arrays, we can successfully recover the spectrum of magnetic turbulence. It is known that interferometer, such as the SKA, combines the signals received from thousands of small antennas to simulate a single giant radio telescope. Thus, for our purpose, the sparsely sampled interferometric data is enough to obtain the power spectrum of turbulent magnetic fields.

\section{Discussion}
It has been demonstrated that important information of MHD turbulence can be extracted from synchrotron polarized radiation. This work employed the method of the power spectrum to carry out our investigations. From a spatial point of view, this method is of two-point statistics. Besides, a variance of the synchrotron polarization intensity, which is a special case of the correlation function at $R = 0$ and called a one-point statistics, can also recover well the slope of turbulence spectrum (LP16, \citealt{Zhang16}). These approaches are complementary each other for exploring the properties of MHD turbulence.

The statistical studies of the intrinsic synchrotron polarization intensity without a wavelength-dependence from Faraday rotation (see  \citealt{Zhang16,Lee16}) are similar to studies of the synchrotron intensity $I$ as done in LP12 and \cite{Herron16}. Compared to statistics of the synchrotron intensity, the advantage of using polarized synchrotron statistics with Faraday rotation is that one can obtain additional information related to the wavelength, which makes statistical measure much more informative. Our present studies focus on recovering the spectral properties of MHD turbulence, whereby we have an insight into fluctuations of anisotropic turbulent magnetic field, electron density, and Faraday rotation density.

We have paid more attention to studying the influence of mean magnetic field on the ring-integrated 1D spectra of $P=Q+iU$ for different spatial configurations in Sections \ref{TESTPERD}, \ref{SSSS} and \ref{SSCS}. With varying the wavelength, the power spectra calculated in different spatial configurations present different shapes. By comparing shapes of the power spectrum at different wavelengths, one can learn about where depolarization occurs, i.e., synchrotron emission source or external Faraday rotation region. It is emphasized that in the foreground medium any new radiation production is not considered for simplicity.

In this paper, we have focused on recovering statistics of synchrotron polarization intensity, $P=Q+iU$. As pointed out in LP16, the synchrotron polarization intensity derivative, $dP/d\lambda^2$, with respect to the square of wavelength, $\lambda^2$, can also used to extract fluctuation information on MHD turbulence, which has been investigated in Appendix A in the case of the spatially separated SEFR regions.

It is very advantageous to adopt the synergy of different techniques to study MHD turbulence. The current work is complementary to the following techniques. The measurements of the kurtosis, skewness as well as genus of gradients of synchrotron polarization radiation, and of multipole moments of normalized structure function of synchrotron intensity can be applied to understand the Mach number of MHD turbulence (\citealt{Gaensler11,Burkhart12,LP16,Herron16}). Very recently, new techniques of both synchrotron intensity gradient (\citealt{Lazarian17}) and synchrotron polarization gradient (\citealt{Lazarian18b}) were proposed to probe 2D and 3D distributions of astrophysical magnetic fields, in which the term `polarization gradient' that is rotationally and translationally invariant is first derived in \cite{Gaensler11}. Furthermore, new development concerning advanced diagnostics of the study of polarization radiation has been done in \cite{Herron18a,Herron18b}. In addition, the velocity channel analysis (VCA) and velocity coordinate spectrum (VCS) can be used to recover statistics of velocity fluctuations (\citealt{LP00,LP04,LP06,LP08}), which is successfully testified and applied to understand observations (\citealt{Laz01,Padoan06,Padoan09,ChepL09,Chep10,ChepL15}). The analysis of moments of the density probabilities (\citealt{Kowal07,Burkhart09,Burkhart10}), Tsallis statistics (\citealt{EL10,Toff11}), bispectra and genus (\citealt{L99,Chep08,Burkhart09}) can be used to get the statistical properties of turbulent density.

In this work, we neglected the influence of polarization synchrotron self-absorption on the power spectrum. It is well known that Faraday rotation would take effect in the relatively long wavelength region. The self-absorption effect should be important for some astrophysical environments and thus considered in the future. Besides, this paper did not consider the correlation between magnetic fields and thermal electron density in synchrotron emission and Faraday rotation regions. The most of the paper is focused on studying how to obtain the spectral properties of magnetic turbulence, such as the perpendicular component, and the line-of-sight (parallel) component of turbulent magnetic field. As done in Sections \ref{SSSS} and \ref{SSCS}, the influence of electron density on the power spectrum is also explored by extracting fluctuations of the Faraday rotation density.

\section{Summary}
We have employed synthetic observations to simulate the power spectra of synchrotron polarization intensity in the spatially coincident, separated, and compounded SEFR regions. As for the former, our numerical results are in agreement with the analytical prediction provided in Equation (103) of LP16. For the two latter, the obtained ring-integrated power spectrum of polarization intensity is slightly more complicated, which has been advanced a bit more beyond what is in LP16. Our results have demonstrated that analytical predictions of the PSA technique in LP16 can be applied to more complicated settings. The numerical results we have obtained are briefly summarized as follows.

1. At very short wavelengths, the ring-integrated 1D spectra reflect fluctuations of the perpendicular component of turbulent magnetic fields. With increasing the wavelength, the shapes of power spectra would be changed due to depolarization of Faraday rotation fluctuation. At long wavelengths, spectra reveal the fluctuation of the Faraday rotation density (see Table \ref{Table:SUM}).

2. At the $k<2\pi\lambda^2\sigma_{\phi}$ part of the wavenumber, Faraday depolarization is significant. On the contrary, the depolarization is less significant in the $k>2\pi\lambda^2\sigma_{\phi}$ regime, from which we can determine the slope of the power spectra.

3. The contribution of the perpendicular component of non-zero mean magnetic field makes amplitudes of the power spectra increase, while its parallel component makes the power decrease due to a lower level of magnetic fluctuations.

4. We could distinguish the origin of magnetic fluctuations from the background synchrotron emission or foreground Faraday rotation region, by comparing the ring-integrated 1D spectra at different wavelengths.

5. Considering effects of telescope's angular resolution and inevitable random noise, we find that the wavenumber $k$ ranging from $k\gtrsim15$ to 100 is an optimal range to determine the spectral slope of MHD turbulence. Therefore, the angular resolution and noise do not present an obstacle for the recovery of the underlying spectra of MHD turbulence.

6. By mimicking interferometric observations, we find that a finite number of baseline arrays would be able to recover the spectral slope of the power spectra.

\acknowledgments  We cordially thank Siyao Xu for her numerous comments that improved the paper. Discussions with Dmitri Pogosyan are also acknowledged. J.F.Z. thanks the supports from the National Natural Science Foundation of China (grant No. 11703020), the CAS Open Research Program of Key Laboratory for the Structure and Evolution of Celestial Objects (grant No. OP201712), and the Hunan Provincial Natural Science Foundation (grant No. 2018JJ3484).  A.L. acknowledges the support of NSF grant DMS 1622353 and AST 1715754.  F.Y.X. thanks the support from the National Natural Science Foundation of China (grant Nos. U1531108 and U1731106).

\appendix

\section{Statistics of synchrotron polarization derivative with respect to $\lambda^2$}
The main part of our paper is focused on the statistics of synchrotron polarization intensity, $P=Q+iU$. In this section, we explore the statistics of synchrotron polarization intensity derivative, $dP/d\lambda^2$, with respect to the square of wavelength, $\lambda^2$, in the case of the spatially separated SEFR regions.

\begin{figure}[]
\centerline{\includegraphics[width=80mm,height=95mm,angle=90]{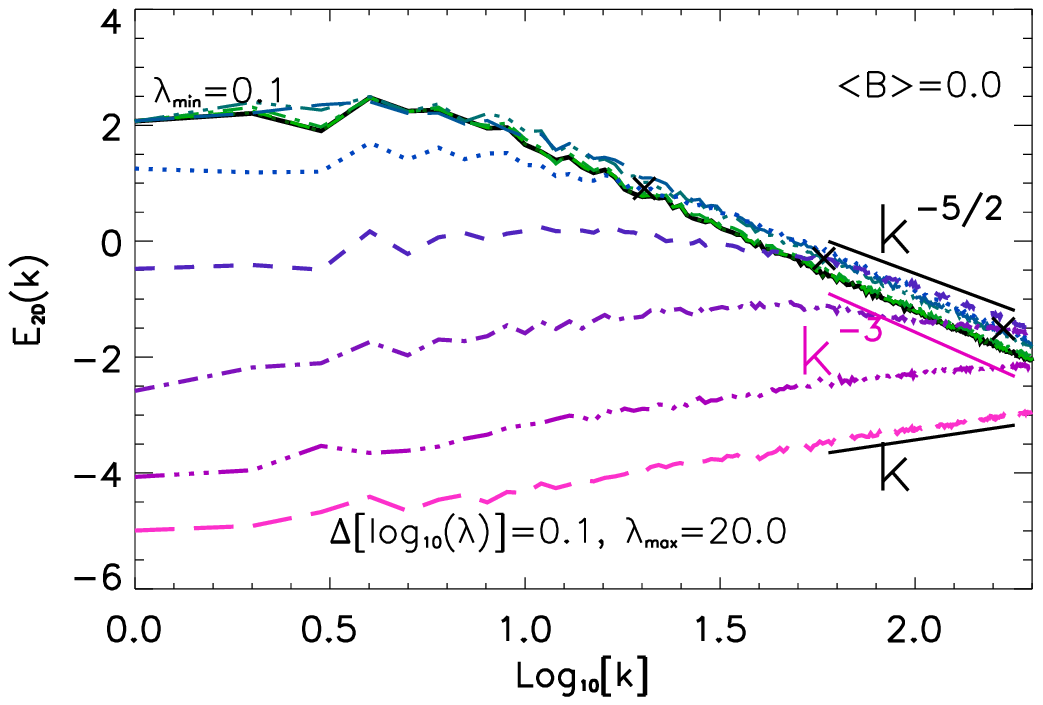} \includegraphics[width=80mm,height=95mm,angle=90]{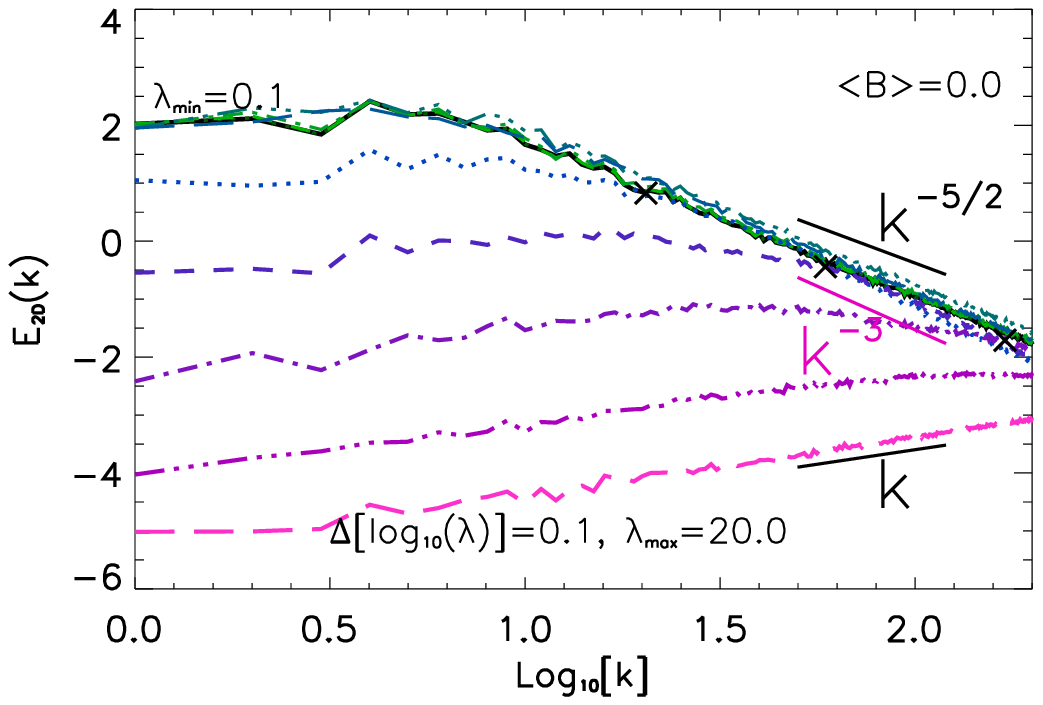}}
\caption{Ring-integrated 1D spectra $E_{\rm 2D}(k)$ of $dP/d\lambda^2$, \emph{in the case of the spatially separated SEFR regions}. The other descriptions are the same as those of Figure \ref{figs:DIFFFRONT}.
 }  \label{figs:DIFFFRONT-dPdl}
\end{figure}

Figure \ref{figs:DIFFFRONT-dPdl} presents the ring-integrated 1D spectra $E_{\rm 2D}(k)$ of $dP/d\lambda^2$ that is calculated using those synthetic data cubes employed in Figure \ref{figs:DIFFFRONT}. It can be seen that the power spectrum of $dP/d\lambda^2$ can recover the spectral properties of MHD turbulence. In the short wavelength range, power spectra reflect the statistics of intrinsic synchrotron polarization fluctuations originating from the perpendicular component of magnetic field, with the scaling slope close to $3$ for the left panel (5/2 for the right panel). With increasing the wavelength $\lambda$, the statistical properties of Faraday rotation density that has a slope close to 5/2 for the left panel (3 for the right panel), can be extracted. This result is consistent with the theoretical prediction provided in Equation (105) of LP16. Compared with the statistics of the polarization intensity $P$, it seems that statistics of $dP/d\lambda^2$ present a less significant transition (by eye) from dominated $B_{\perp}$ fluctuation to dominated fluctuations of Faraday rotation density $\phi$. If one would like to distinguish shape of the power spectrum at different wavelengths, the statistics of $P$ should be preferred, because the derivative procedure for a polarization map, with respect to $\lambda^2$, may smooth some small statistical features.

\end{document}